\documentclass[useAMS,usenatbib,usedcolumn]{mn2e}

\pdfoutput=1
\usepackage{graphicx}
\usepackage{xcolor}
\usepackage{booktabs}
\newcommand{\ra}[1]{\renewcommand{\arraystretch}{#1}}

\usepackage{flushend}

\title[Structure formation with suppressed perturbations]{Structure formation with suppressed small-scale perturbations}
\author[Aurel Schneider]{Aurel Schneider\\
{Center for Theoretical Astrophysics and Cosmology, Institute for Computational Sciences, University of Zurich, Switzerland}\\
{Department of Physics and Astronomy, University of Sussex, Brighton, United Kingdom}\\
{Email: aurel@physik.uzh.ch}}

\begin{document}

\label{firstpage}
\maketitle

\begin{abstract}
All commonly considered dark matter scenarios are based on hypothetical particles with small but non-zero thermal velocities and tiny interaction cross-sections. A generic consequence of these attributes is the suppression of small-scale matter perturbations either due to free-streaming or due to interactions with the primordial plasma. The suppression scale can vary over many orders of magnitude depending on particle candidate and production mechanism in the early Universe.

While nonlinear structure formation has been explored in great detail well above the suppression scale, the range around suppressed perturbations is still poorly understood. In this paper we study structure formation in the regime of suppressed perturbations using both analytical techniques and numerical simulations. We develop simple and theoretically motivated recipes for the halo {\it mass function}, the expected {\it number of satellites}, and the halo {\it concentrations}, which are designed to work for power spectra with suppression at arbitrary scale and of arbitrary shape. As case studies, we explore warm and mixed dark matter scenarios where effects are most distinctive. Additionally, we examine the standard dark matter scenario based on weakly interacting massive particles (WIMP) and compare it to pure cold dark matter with zero primordial temperature. We find that our analytically motivated recipes are in good agreement with simulations for all investigated dark matter scenarios, and we therefore conclude that they can be used for generic cases with arbitrarily suppressed small-scale perturbations.
\end{abstract}



\section{Introduction}
Within the last decade, numerical simulations have substantially increased in performance and accuracy contributing to what has been denominated the new era of precision cosmology. Statistical measures such as the matter power spectrum can now be calculated to sub-percent level precision, making it possible to obtain competitive constraints of fundamental cosmological parameters with galaxy surveys. This precision is, however, only obtained in a regime where structure formation behaves in a regular, strictly hierarchical way, owing to a well behaved linear power spectrum with a close to power-law scaling. As soon as this basic assumption is relaxed, standard numerical methods are not guaranteed to work anymore. In the case of a suppressed power spectrum, as it appears prominently in hot and warm dark matter models, standard numerical methods produce artefacts near the suppression scale by enhancing small non-physical modes present in the initial conditions \citep{Goetz2003,Wang2007}.

Along with the development of numerical schemes, analytical techniques such as the extended Press-Schechter (EPS) approach \citep{Press1974, Bond1991} and the halo model \citep{Cooray2002} have been put forward to describe nonlinear structure formation. They do not achieve the same precision than numerical simulations, but they provide qualitative understanding of the physical processes involved and they have the advantage of not requiring big computer facilities. However, these analytical techniques again only work in the case of a reasonably behaved initial power spectrum with nearly power-law scaling. As soon as the initial power spectrum is significantly suppressed, standard EPS approaches and subsequently the halo model fail to predict the right clustering \citep[][]{Schneider2013}.

Understanding structure formation of cosmologies with suppressed perturbations is not merely an academical exercise but a necessity in order to capture the full range of scales of cosmic clustering. As a matter of fact, all reasonable dark matter (DM) scenarios exhibit a suppression of perturbations below a certain scale either due to particle free-streaming or due to interactions in the early Universe. For the prime DM candidate, the weakly interacting massive particle (WIMP), the mass scale of suppression lies somewhere between roughly $10^{-12}$ $h^{-1}\rm M_{\odot}$ and $10^{-4}$ $h^{-1}\rm M_{\odot}$ depending on the specific model parameters \citep{Green2005, Profumo2005}, which is orders of magnitude smaller than the smallest galaxies we observe\footnote{The only chance of observing such small structures is probably by directly measuring the flux from dark matter annihilation.}. Other promising DM candidates such as the sterile neutrino or the gravitino have higher primordial velocities, pushing the regime of power suppression to larger scales where it becomes relevant for galaxy surveys such as in the case of warm or mixed DM. More exotic scenarios like interacting DM \citep{Boehm2005}, self-interacting DM \citep{Spergel2000}, asymmetric DM \citep{Petraki2013}, or ultra-light axion DM \citep{Marsh2013} can also lead to suppressed power at rather large scales either because they interact with the cosmic plasma, with dark radiation or because they undergo scalar field oscillations on astrophysical scales. Since the power suppression of different DM scenarios happens at different scales and is of varying shape, it is crucial to properly quantify the nonlinear clustering in this regime in order to distinguish between different DM species and to contribute towards a solution of one of the outstanding puzzles in modern physics.

In this paper we study nonlinear structure formation starting from initial power spectra with arbitrary small-scale suppressions. As working examples, we investigate the cases of warm DM, mixed DM, WIMP DM, and pure cold DM. We run a suite of $N$-body simulations for these models and carefully remove artificial structures which tend to populate simulations of suppressed initial power. At the same time, we develop an extended Press-Schechter method able to cope with arbitrary linear power spectra. We then calculate important quantities, such as the halo mass function, the number of satellites, and the concentration-mass relation of halo profiles.

The paper is structured as follows: In Sec. \ref{sec:DMmodels} we give a short overview of different DM scenarios and discuss how they suppress perturbations. Sec. \ref{sec:simulations} contains a summary of the simulations and analysis techniques including a discussion about subtracting artefacts. In Sec. \ref{sec:EPS} and \ref{sec:concentrations} we present our modified Press-Schechter approach and use it to predict mass function, number of Milky-Way satellites and concentrations. Finally, we conclude in Sec. \ref{sec:conclusions} and give further details about the removal of artefacts in the Appendix.


\section{Dark Matter Models and the Suppression of Perturbations}\label{sec:DMmodels}
The physical mechanisms leading to the power suppression below a certain scale depend on the dark matter (DM) candidate but is usually either particle free-streaming, tiny interactions with the primordial plasma, or more exotic phenomena in the early Universe. Both the scale and the detailed shape of the suppression strongly vary depending on the characteristics of the DM particle, i.e. its mass, momentum, and interaction cross-sections. In the following, we discuss the most common DM scenarios and outline their effects on nonlinear structure formation.

\begin{figure*}
\centering
\includegraphics[trim = 6mm 10mm 42mm 12mm, clip, scale=0.42]{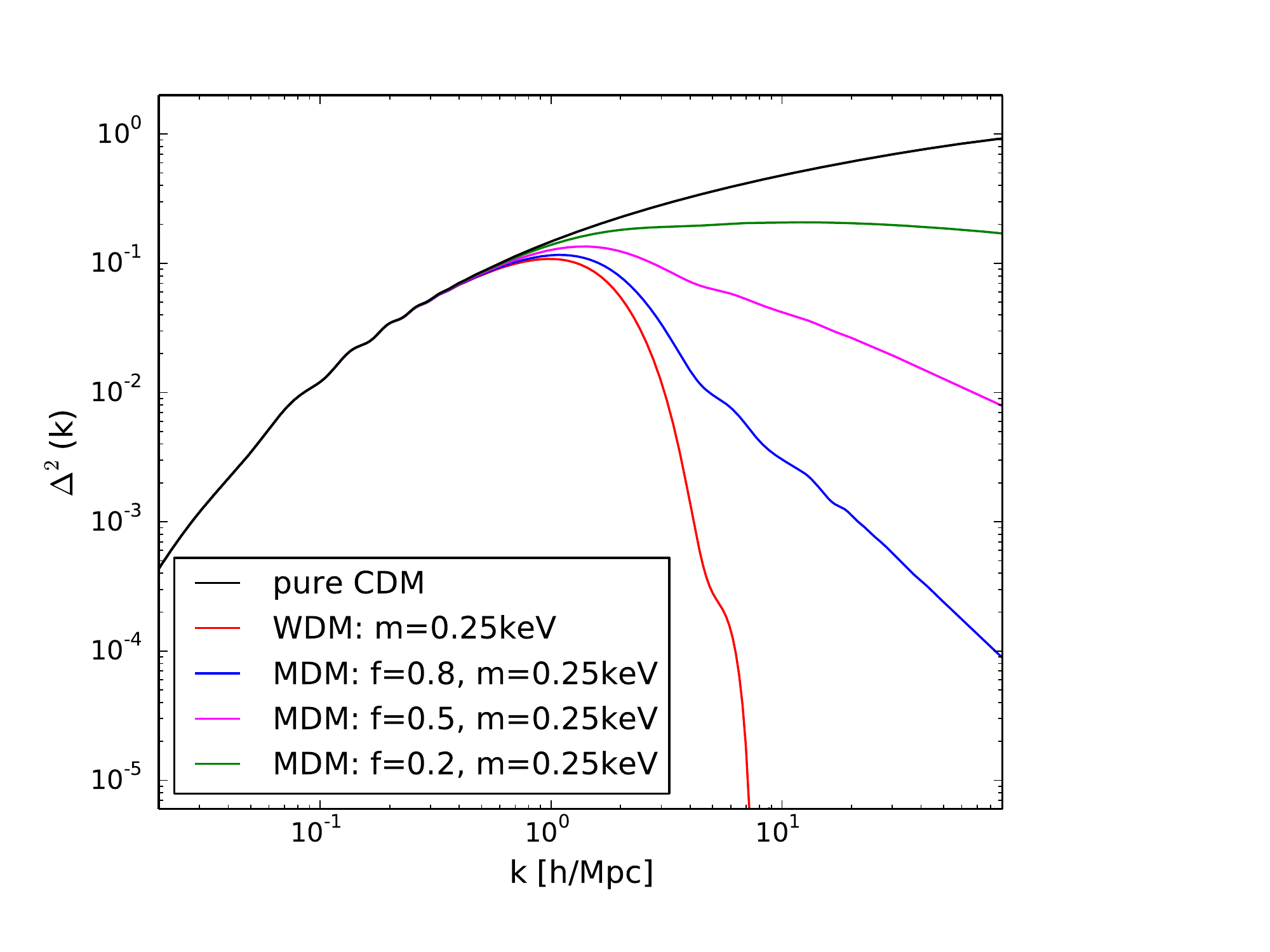}
\includegraphics[trim = 6mm 10mm 42mm 12mm, clip, scale=0.42]{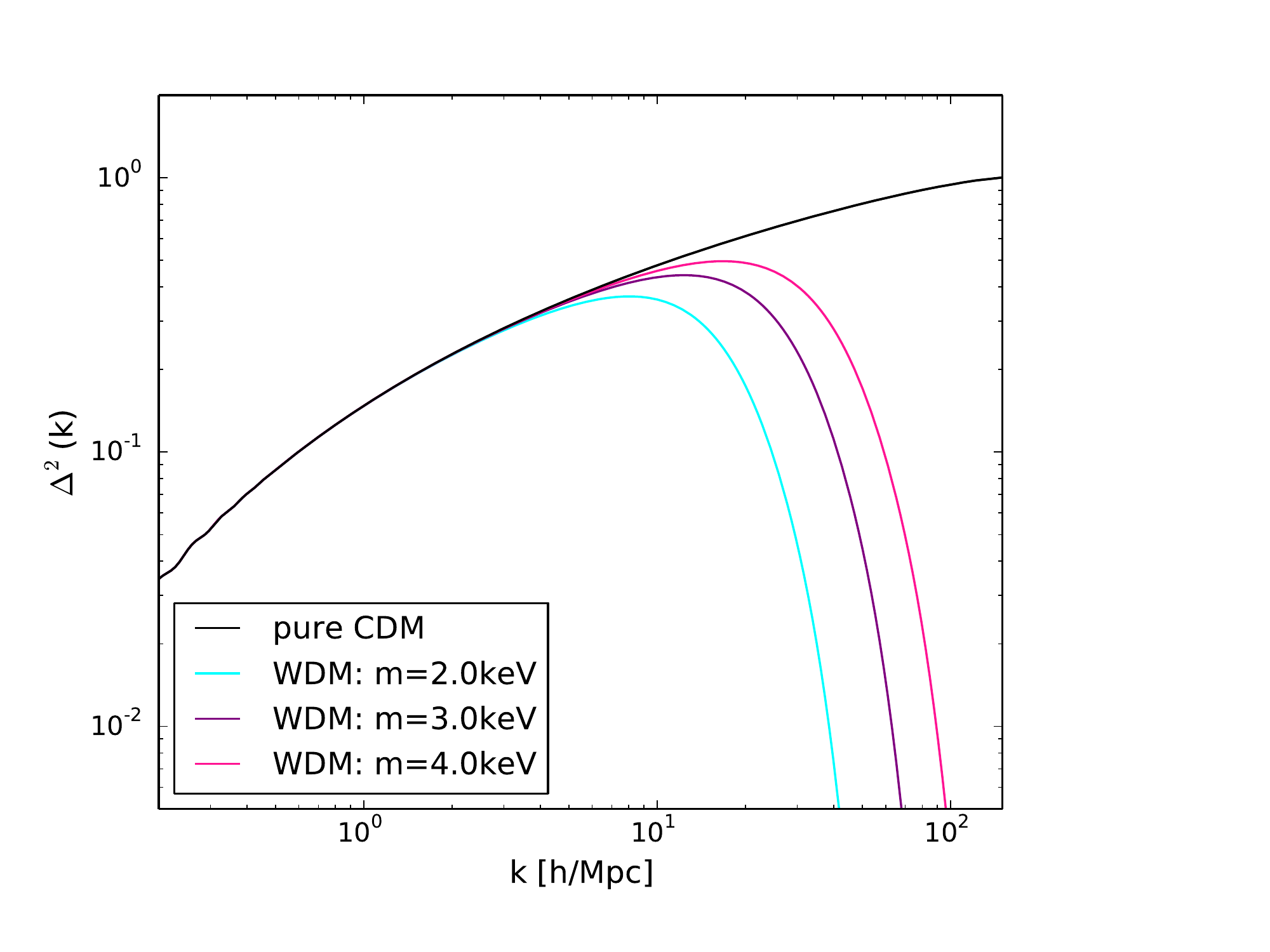}
\includegraphics[trim = 6mm 10mm 98mm 12mm, clip, scale=0.42]{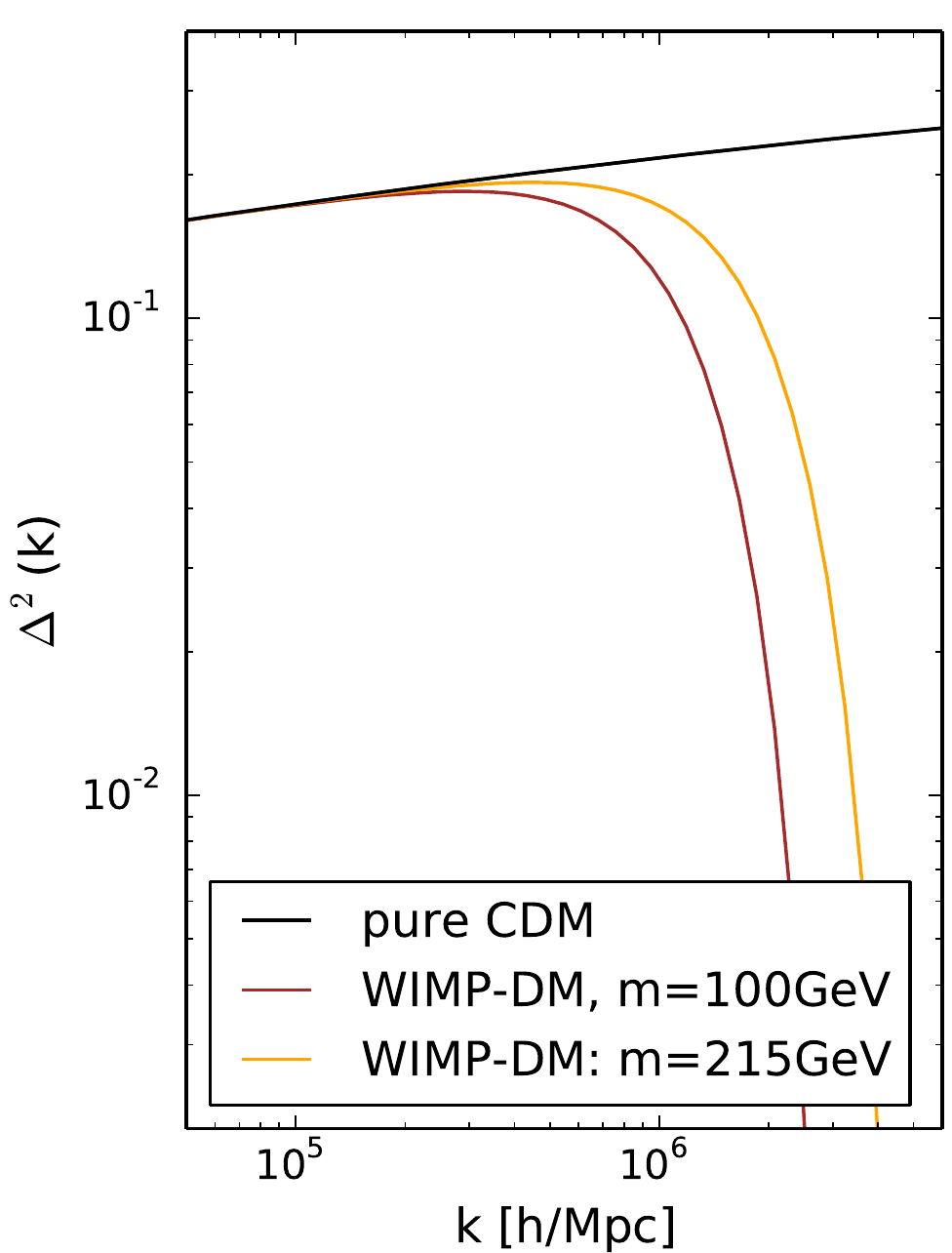}
\caption{Linear dimensionless power spectrum $\Delta(k)=k^3P_{\chi}(k)/2\pi^2$ of the dark matter scenarios ($\chi$) investigated in this paper. Left: CDM (black), WDM (red) and various MDM models (blue, magenta, green) at redshift 50. Middle: CDM (black) and different WDM models (cyan, purple, pink) at redshift 100. Right: pure CDM (black) and two WIMP-DM scenarios (brown, orange) at redshift 300.}\label{fig:powerspectrum}
\end{figure*}

\subsection{Warm dark matter (WDM)}
Dark matter models with a steep cutoff-suppression at dwarf galaxy scales, similar in shape to the one obtained from particles with a Fermi-Dirac momentum distribution, are usually referred to as warm DM (WDM). Typical power spectra of WDM models are illustrated in Fig. \ref{fig:powerspectrum} (red, cyan, purple, and pink lines). Implicitly, the WDM regime is defined to lie in the rather narrow band where the suppression of perturbation has a significant effect on dwarf galaxy formation but is still in agreement with observations. Furthermore, WDM is generally assumed to reduce potential small-scale inconsistencies, such as the {\it missing satellite} or the {\it too big to fail} problem, occurring in a $\Lambda$CDM universe \citep[see][for a review on these topics]{Weinberg2013}. 

Recent studies point out that both requirements -- passing observational constraints and alleviating small scale inconsistencies -- seem impossible to combine because constraints from Lyman-$\alpha$ are prohibitively strong in the case of WDM \citep{Viel2013, Schneider2014}. While some question the accuracy of the Lyman-$\alpha$ measurements, more and more people believe that the small-scale inconsistencies are a consequence of poorly understood baryonic processes of galaxy formation \citep{Brooks2014}.

However, it is possible and worthwhile to use structure formation in order to find upper limits on the scale of DM suppression. The strongest constraints currently come from flux observations of the Lyman-$\alpha$ forest ruling out particle masses below $m_{\rm TH}=3.3$ keV at the 2-$\sigma$ confidence level \citep{Viel2013}. Other limits are obtained from dwarf galaxy counts disfavouring particle masses below $m_{\rm TH}=2.3$ keV \citep{Polisensky2011,Kennedy2014}. Here $m_{\rm TH}$ refers to the equivalent thermal mass, assuming a Fermi-Dirac momentum distribution.

The prime candidate for WDM is the hypothetical sterile (or right-handed) neutrino, which can be readily added to the standard model. Sterile neutrinos are well motivated because they provide an explanation for the measured mixing angles of active neutrinos, and because all other fermions exist with both left and right chirality \citep{Drewes2013}. The particle mass of sterile neutrino DM is expected to lie in the keV-range, what makes it an ideal candidate for warm or lukewarm dark matter depending on the suppression scale which is governed by particle mass and momentum distribution.

The recent tentative discovery of an X-ray line at 3.5 keV from galaxy clusters and M31 by two independent research groups \citep{Bulbul2014, Boyarsky2014} might provide additional motivation for sterile neutrino DM. Since sterile neutrinos decay radiatively, the confirmation of such a signal would be a smoking gun for a $m_{\rm sn}=7.1$ keV particle, which would translate into a thermal mass of $m\simeq 2.5 - 5$ keV depending on the production mechanism in the early Universe \citep{Abazajian2014, Merle2014}. The measured X-ray line is, however, disputed, since there are some conflicting bounds from galaxies where the signal is not measured \citep[see][for a review]{Iakubovskyi2014}. Better astronomical data is needed to check if the signal is real and if the flux coincides with the expectations from decaying DM.

\subsection{Mixed dark matter (MDM)}
In principle, dark matter can be made of two or more particle species with different properties. Particularly interesting in terms of structure formation is a mixture of cold with warm or hot DM. This leads to qualitatively different initial power spectra with gradual suppression over many orders of magnitude in scale in contrary to the steep cutoff known from WDM. The effect is illustrated in the left panel of Fig. \ref{fig:powerspectrum}, where we plot the dimensionless power spectra of different mixed DM (MDM) scenarios (consisting of a CDM part and a thermal WDM part with $m=0.25$ keV) with increasing mixing fraction $f=\Omega_{\rm WDM}/(\Omega_{\rm CDM}+\Omega_{\rm WDM})$. 

The concept of MDM is neither new nor particularly exotic. In fact, it is clear that there must be more than one DM component since neutrinos are known to have non-zero mass. Current observations constrain the sum of neutrino masses to the range of $0.05-0,23$ eV \citep{Fogli2012,Planck2014}, yielding a mixing fraction $f\sim0.004-0.02$. However, the very small neutrino masses result in an evenly damped power spectrum at all relevant scales for halo formation, very similar to the case of pure CDM with low $\sigma_8$-normalisation \citep{Viel2010}.

Instead of two (or more) distinct particles acting as DM components, MDM-like compositions can also arise due to multi-channel DM production in the early Universe, yielding momentum distributions that mimic the case of several DM components. Prime example is again the sterile neutrino which can be produced via resonant oscillations with active neutrinos \citep{Shi1999}, where some subdominant part is always produced out of resonance leading to particle momenta from two overlapping distributions \citep{Boyarsky2009}. The effect can be even stronger if the sterile neutrinos with nonzero mixing angle are produced via the decay of heavy scalars yielding a momentum distribution with two distinct peaks \citep{Merle2014,Merle2015}.

\subsection{WIMP dark matter}
The most popular group of dark matter candidates are weakly interacting massive particles (WIMP) with the neutralino as prime candidate. The popularity of WIMP DM comes from the fact that such particles naturally appear in supersymmetric extensions of the standard model, and that they are produced via thermal freeze-out at roughly the right amount to account for the observed DM abundance \citep[this is usually referred to as the {\it WIMP miracle}, see][for a summary]{Bertone2005}.

As WIMPs are heavy (with particle masses in the GeV or TeV scales) and weakly interacting, they become non-relativistic very early leading to extremely small suppression scales. Depending on the parameters of the model, the mass scale of WIMP DM suppression is expected to lie between roughly $10^{-12}$ $h^{-1}\rm M_{\odot}$ and $10^{-4}$ $h^{-1}\rm M_{\odot}$ \citep{Profumo2005}.

In the right panel of Fig.~\ref{fig:powerspectrum} we plot the power spectra of neutralino DM with a mass of $m=100$ (brown line) and $m=215$ GeV (orange line) and corresponding decoupling temperatures of $T_{\rm dk}=28$ MeV and $T_{\rm dk}=33$ MeV. These spectra are compared to the hypothetical case of pure CDM (black line). The suppression of power in WIMP scenarios happens at very high wave-numbers roughly following an exponential cutoff \citep{Green2005}. We will show later on that such WIMP models suppress structure formation at a halo mass of about $10^{-6}$ $h^{-1}\rm M_{\odot}$.

\subsection{Other models with suppressed power}
There are many more DM candidates with variable suppression scales depending on their interaction and free streaming properties. For example, DM can be self-interacting, as in the case of atomic or mirror DM \citep{Cyr-Racine2013}, or it can interact with photons \citep{Boehm2014,Wilkinson2014a} and neutrinos \citep{Wilkinson2014b} all yielding strong suppression of small-scale modes. Other possibilities with similar effects on small scales are decaying DM \citep{Kaplinghat2005,Wang2014}, later-forming DM \citep{Agarwal2014}, or ultra-light axion DM \citep{Marsh2013}.

It is also possible to obtain suppressed small-scale perturbations from effects not related to dark matter. Inflation could lead to a running of the spectral index, gradually reducing power on small scales \citep{Kosowsky1995}, or it could induce a strong cutoff similarly to the case of WDM \citep{Kamionkowski2000}.


\section{Numerical simulations}\label{sec:simulations}
We run and analyse numerical simulations of different resolution with linear power spectra representing cold, warm, mixed, and WIMP dark matter (DM) scenarios. The initial conditions are generated from the linear power spectra illustrated in Fig. \ref{fig:powerspectrum}, which are selected to cover different scales of power suppression as well as a variety of shapes from steep cutoffs to shallow decreases towards large wave-numbers.

The linear power spectra have been calculated with the Boltzmann solver {\tt CLASS} \citep{Blas2011}, which comes with an option for multiple DM species of arbitrary mass and momentum distribution \citep{Lesgourgues2011}. We use cosmological parameters obtained by Planck, i.e $\Omega_{\rm m}=0.304$, $\Omega_{\rm b}=0.048$, $\Omega_{\Lambda}=0.696$, $H_0=68.14$, $n_s=0.963$, and $\sigma_8=0.827$ \citep{Planck2014}. The simulations are set up with the initial conditions generator {\tt MUSIC} \citep{Hahn2011} using second order Lagrangian perturbation theory and initial redshifts of $[50, 100, 200, 200, 300]$ for runs with box-size $[256, 64, 8, 4, 0.0001]$ $h^{-1}\rm Mpc$. The simulations are performed with {\tt PKDGRAV2} \citep{Stadel2001}, applying a standard force softening of 1/50 times the mean particle separation.  For the halo finding, we use {\tt AHF} with particle unbinding and an over-density criterion corresponding to spherical tophat collapse \citep{Gill2004,Knollmann2009}. A list summarising the main characteristics of the simulations is given in Table \ref{sims}.

\begin{table}
\ra{1.2}
   \centering{
\begin{tabular}{llll}
  \hline
  Scenario: && $L$ [$h^{-1}\rm Mpc$] & $z_f$ \\
  \hline
  CDM: & pure & 24, 64, 256  & 0\\
  WDM: & $m$=0.25 keV & 64, 256 & 0 \\
  MDM: & $m$=0.25 keV, $f$=0.8 & 64  & 0 \\
  MDM: & $m$=0.25 keV, $f$=0.5  & 64 & 0\\
  MDM: & $m$=0.25 keV, $f$=0.2 & 64  & 0\\ 
  WDM: & $m$=2.0 keV &  24  & 0\\
  WDM: & $m$=3.0 keV &  24  & 0\\
  CDM: & pure & 8  & 5\\
  WDM: & $m$=2.0 keV & 8  & 5\\
  WDM: & $m$=3.0 keV & 8  & 5\\
  WDM: & $m$=4.0 keV & 4  & 5\\
  CDM: & pure & $10^{-4}$ & 23\\
  WIMP: & $m$=100 GeV, $T_{\rm kd}$=28 MeV& $10^{-4}$  & 23\\
  WIMP: & $m$=215 GeV, $T_{\rm kd}$=33 MeV & $10^{-4}$  & 23\\   
  \hline
\end{tabular}}
\caption{List of simulations performed in this paper. 
The parameters $m$, $f$, $T_{\rm kd}$ refer to (thermal) particle mass, mixing ratio, and decoupling temperature.  $L$ [$h^{-1}\rm Mpc$] designates the box-size and $z_f$ the final redshift. All simulations have been run at both low and high resolution (with $512^3$ and $1024^3$ particles, respectively).}\label{sims}
\end{table}

Most of the runs suffer from artificial clumping at the scale of suppressed perturbations. This is not only the case for WDM, where the effect has been observed many times before, but also for WIMP-DM and even for the MDM scenarios with initial power spectra that never drop to zero. In the next sections, we discuss the issue of artificial halo formation in more detail, and we discuss how artefact can be filtered out of halo catalogues from numerical simulations.

\subsection{The problem of artificial clumping}
Artificial clumping is a serious problem in simulations where the initial power is suppressed below a characteristic scale, usually referred to as the free-streaming or damping scale. Standard N-body techniques cannot cope with this setup and produce many small-scale haloes that are resolution dependent and therefore non-physical \citep{Goetz2003, Wang2007}. In this paper we show that the artificial clumping does not only happen in setups with a steep cutoff and zero physical power below a certain scale, but also in MDM scenarios, where the suppression of perturbations is much smoother and the power spectrum never drops to zero.

Recently, there has been attempts to overcome the problem of artefacts with a novel N-body technique tracking the dark matter sheet in phase-space \citep{Abel2012}. With this method, the amount of artificial haloes in WDM simulations is drastically reduced \citep{Angulo2013}. The method is, however, not fully operational yet, since it has difficulties to cope with high-density regions, where the six-dimensional phase-space sheet undergoes multiple foldings. Further studies seem needed in order evaluate whether this promising method can be improved to simultaneously deal with artificial clumping and accurately calculate regions of high density.

In this paper we use a standard N-body technique combined with a post-processing method to identify and remove artificial haloes from the halo catalogue. The method is explained in the following section.

\subsection{Removing artefacts}\label{sec:remart}
Several methods of different complexity to identify artificial haloes have been proposed in the past. \citet{Diemand2005} removed haloes on the basis of visual inspection, \citet{Schneider2013} suggested to subtract a power-law component in the halo mass function, \citet{Agarwal2015} identified artefacts by measuring the spin of haloes found with the friends-of-friends technique. The most complex and arguably most complete method to remove artefacts has been proposed by \citet{Lovell2014}, using both the shape of proto-haloes in the initial conditions (i.e. the volume spanned by all particles ending up in a halo) as well as their overlap between realisations with different resolution. In a first step, \citet{Lovell2014} measured the sphericity of proto-haloes, removing all objects that are unusually elongated. This selection criterion is based on the observation that artefacts originate from the collapse of larger modes to walls and filaments, pushing already aligned particles together until they cluster into non-physical clumps. In a second step, they analysed the resolution dependence of haloes, removing all objects that do not appear in two simulations of the same initial field but different resolution.

\begin{figure*}
\includegraphics[trim = 8mm 4mm 20mm 10mm, clip, scale=0.49]{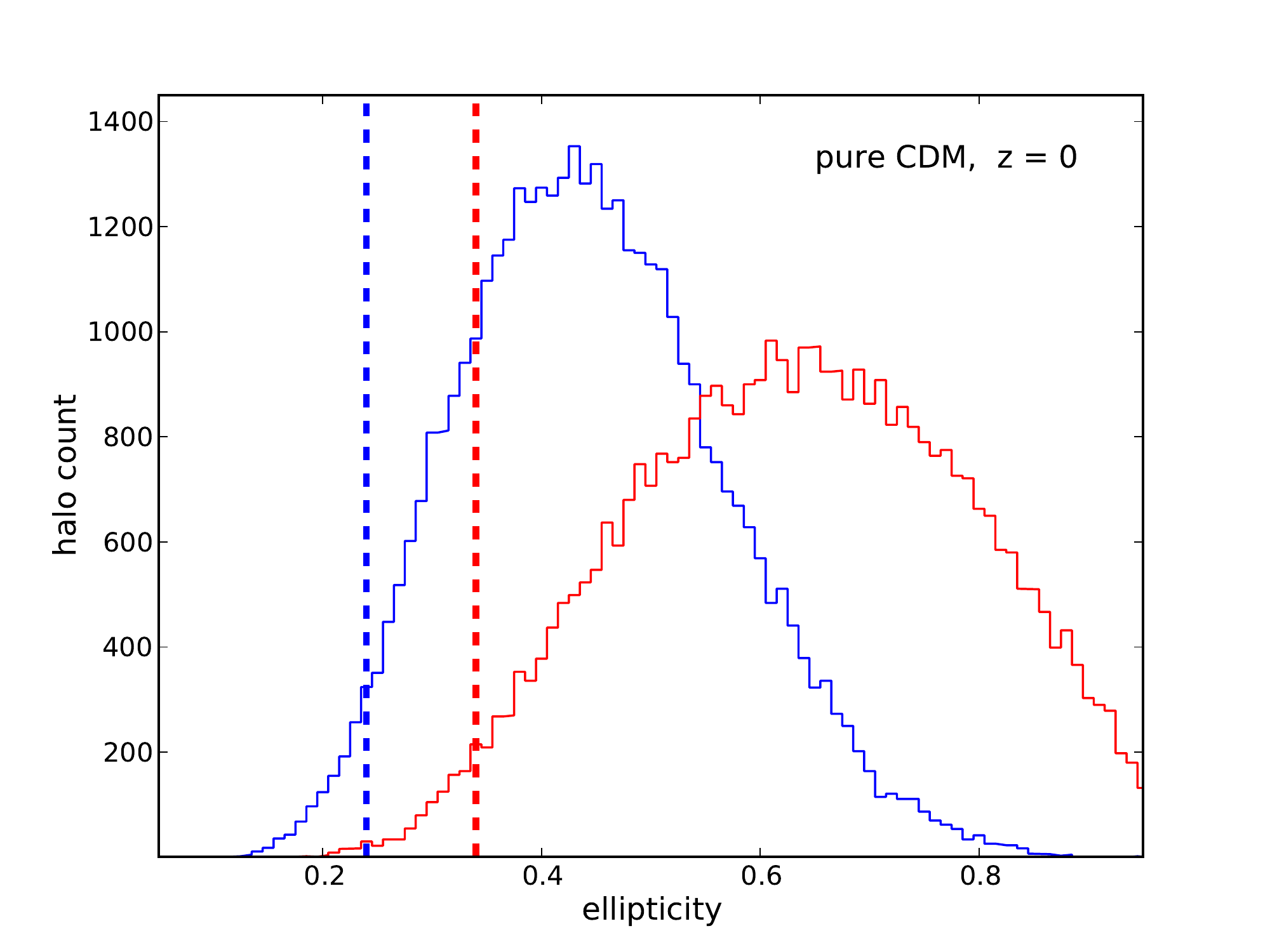}
\includegraphics[trim = 12mm 4mm 20mm 10mm, clip, scale=0.49]{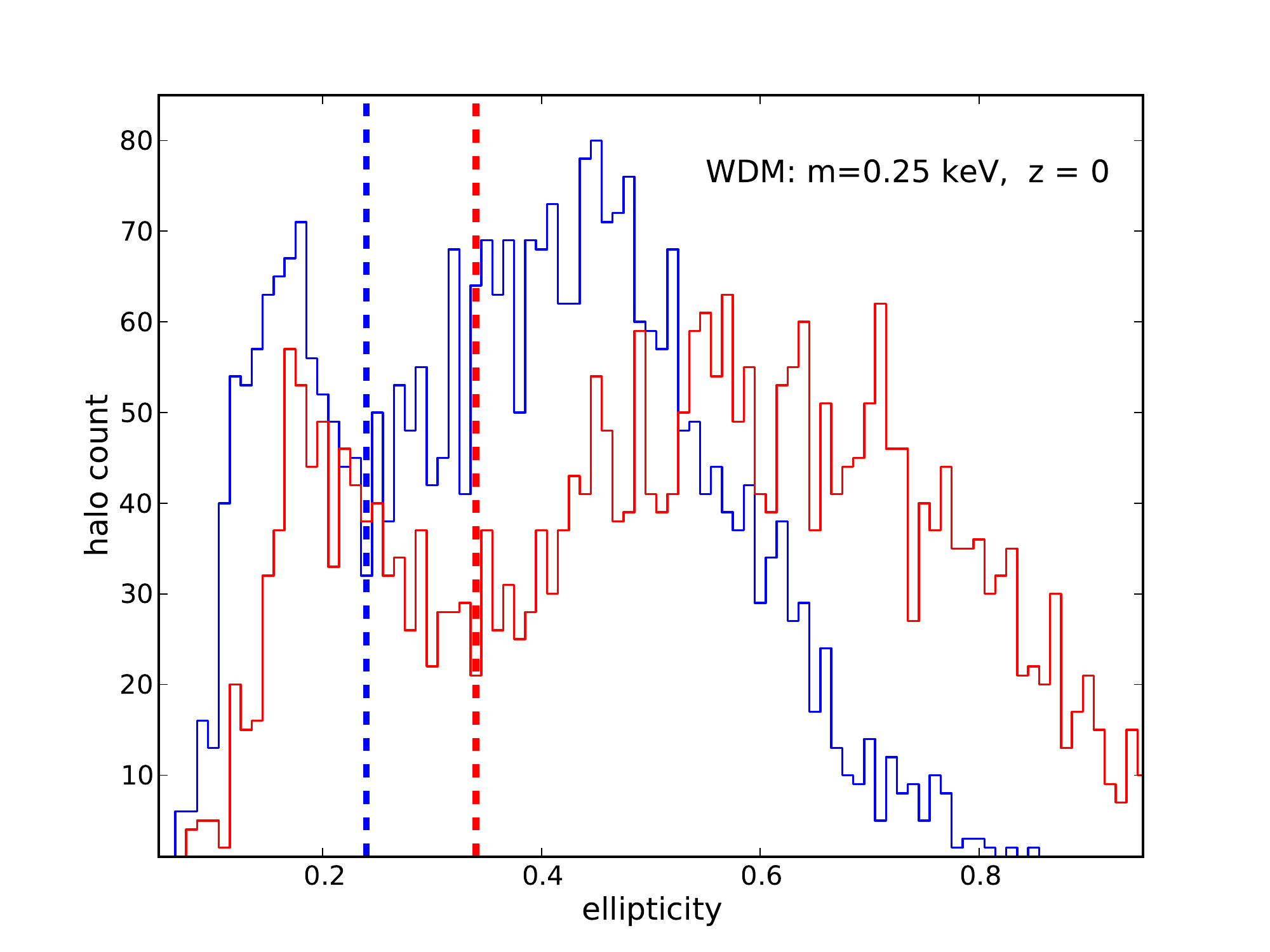}\\
\includegraphics[trim = 0mm 4mm 20mm 10mm, clip, scale=0.312]{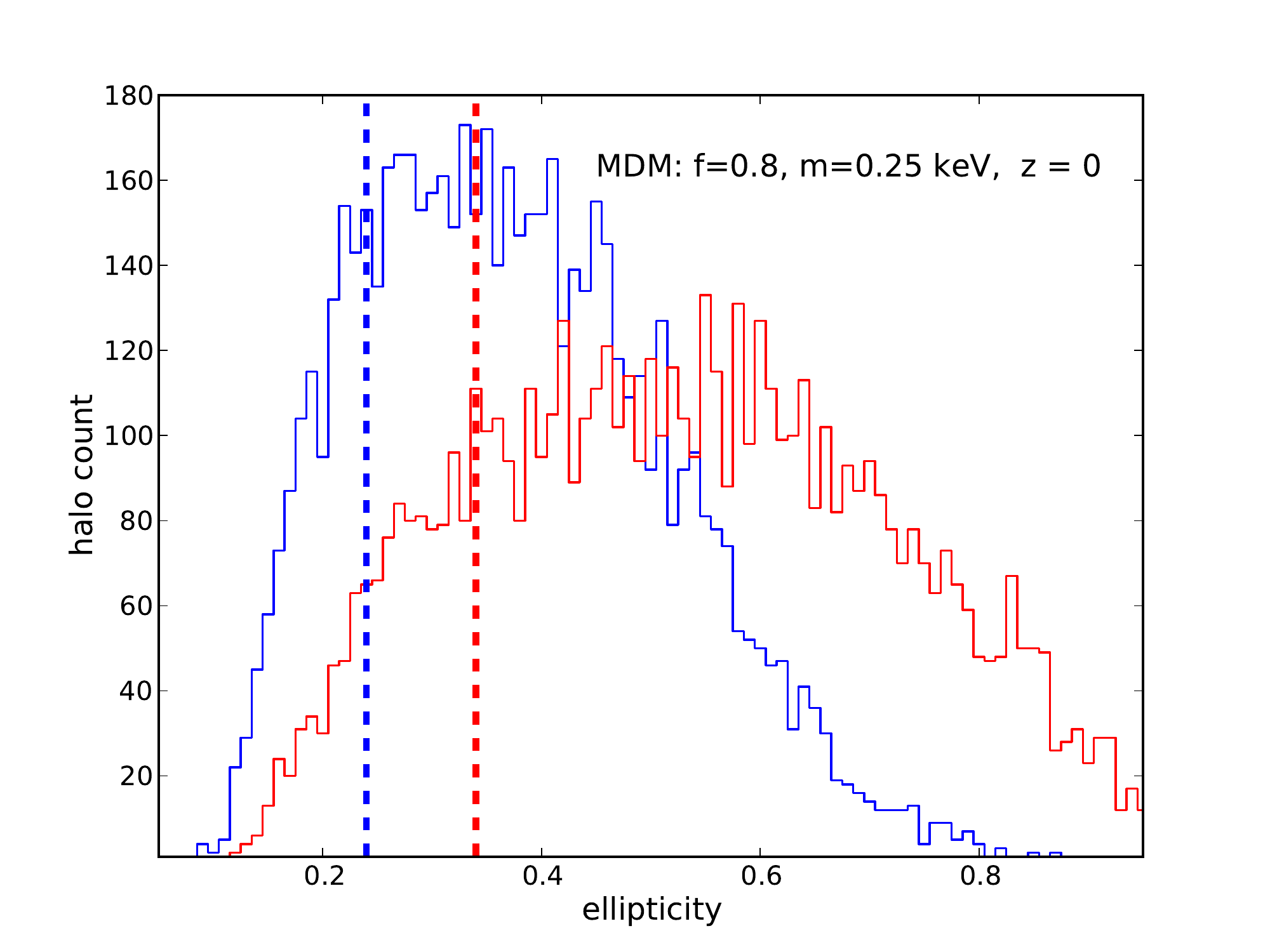}
\includegraphics[trim = 6mm 4mm 20mm 10mm, clip, scale=0.31]{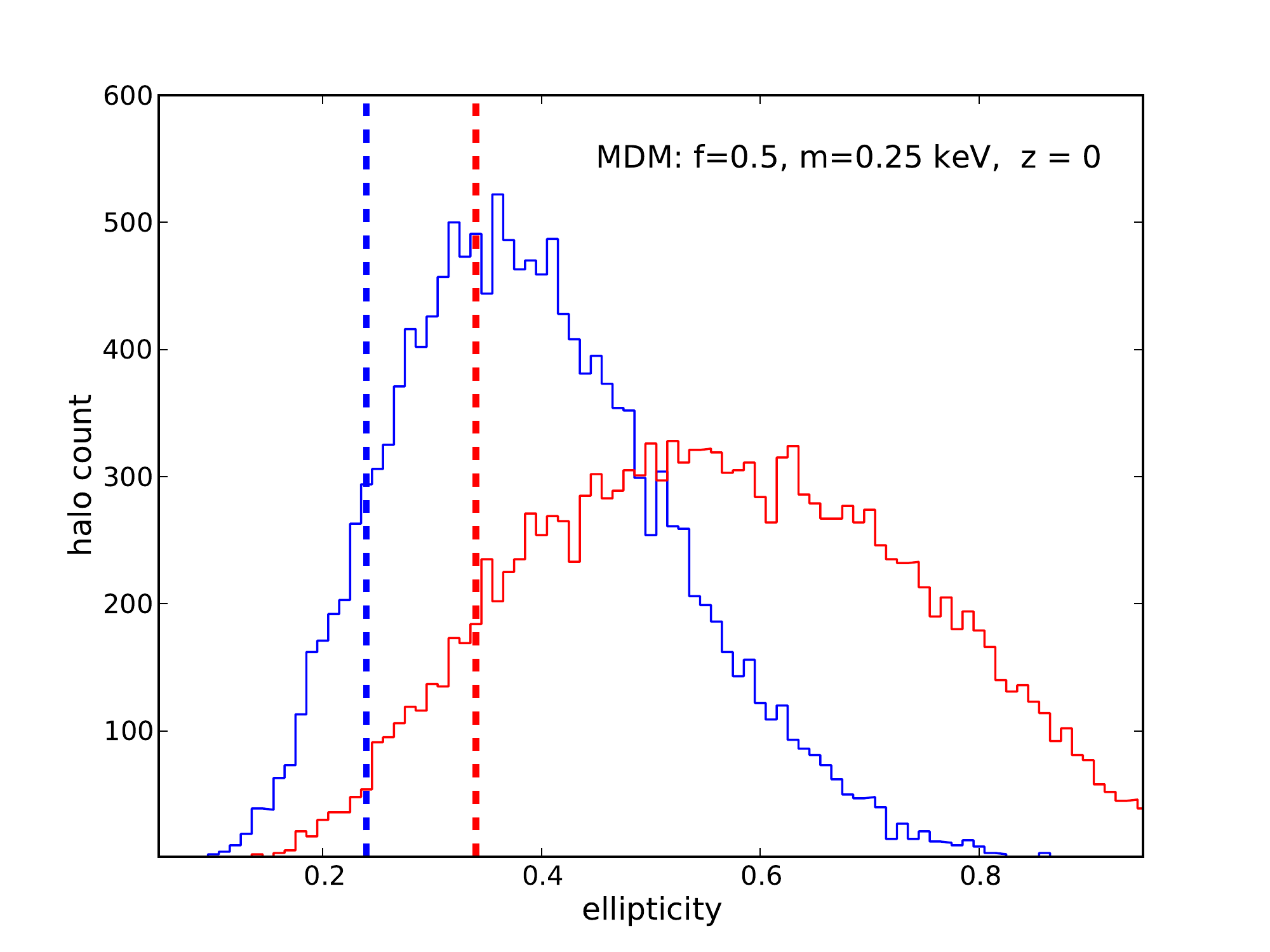}
\includegraphics[trim = 6mm 4mm 20mm 10mm, clip, scale=0.312]{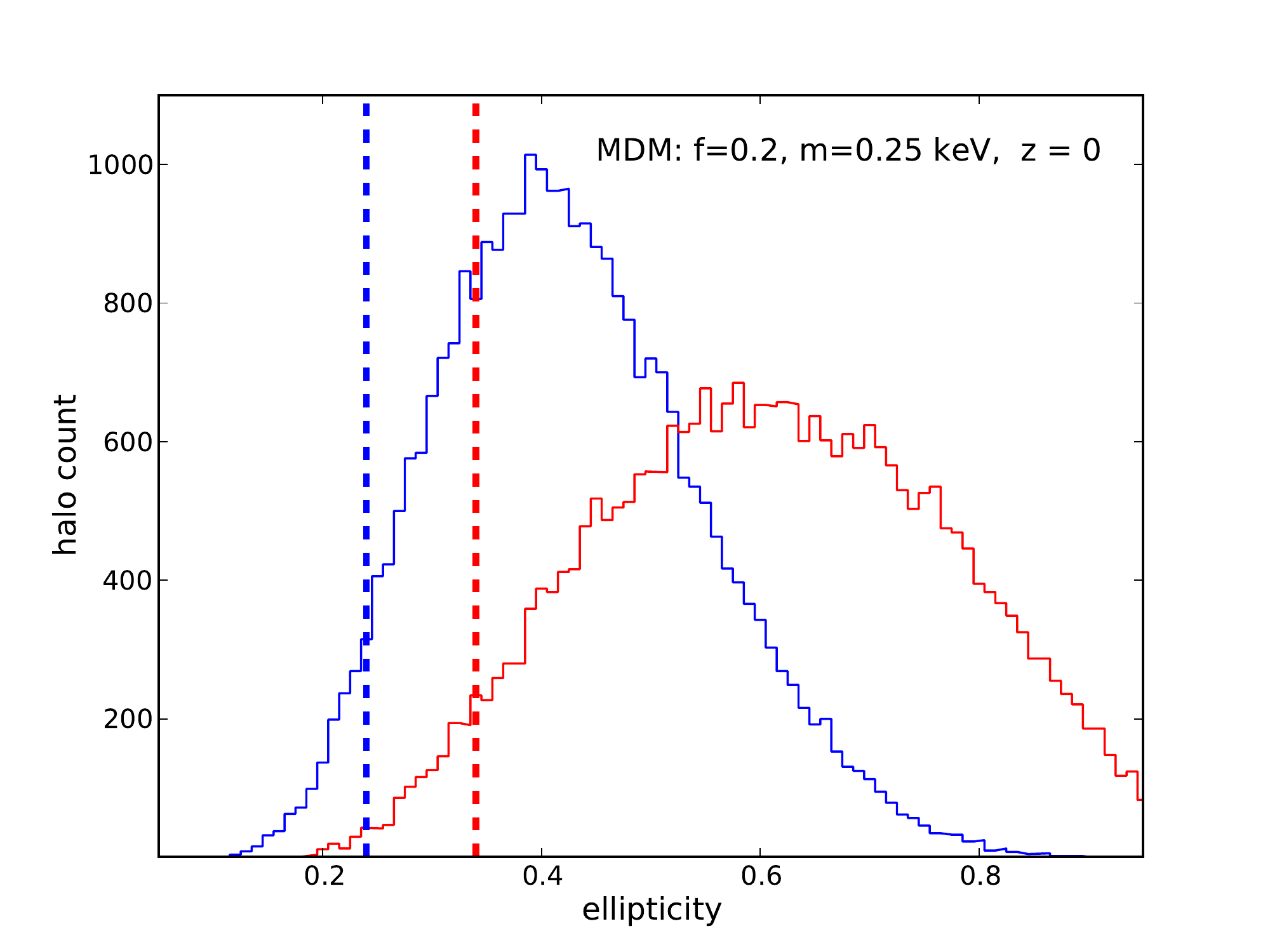}
\caption{Histograms of haloes with respect to the ellipticity parameters $s=c/a$ (blue) and $q=c/b$ (red), where $a\geq b\geq c$ are the semi-major axes of an ellipsoid with the same inertia tensor than the proto-halo. The top row illustrates the cases of CDM (left) and WDM with $m=0.25$ keV (right), the bottom row the cases of MDM with $f=0.8$ (left) $f=0.5$ (middle) and $f=0.2$ (right). Vertical dashed lines at 0.24 and 0.34 designate the {\it sphericity cut} applied to separate real haloes from artefacts.}
\label{fig:halohistogram}
\end{figure*}

In this paper, we use a similar method than the one proposed by \citet{Lovell2014} and apply it to cosmological simulations. This is more challenging than the original application which was limited to substructures within a Milky-Way-size host haloes. In our case we use a less stringent limit of 1000 particles for the minimal halo size, which means that we have up to a quarter of a million haloes per simulation with masses spanning more than four orders of magnitude. These haloes are in varying large-scale surroundings and at different stages of their evolution, making it hard to reliably separate real haloes from artefacts. In the following, we show that although no perfect separation is possible, most artefacts can be reliably identified while only a small fraction of real haloes is erroneously removed. We use the simulations with the fewest and the most artefacts -- the CDM and the WDM ($m=0.25$ keV) run -- to illustrate the method and estimate the efficiency of filtering out artefacts without erroneously removing physical haloes. Further information on how halo catalogues of the remaining DM models are affected by the method is provided in the Appendix. 

The first step of the algorithm consist of measuring the ellipticity of proto-haloes in the initial conditions (ICs) using the inertia tensor
\begin{equation}
I_{ij}=m \sum_{\rm particles} \left(\delta_{ij} \mathbf{x}^2- x_i x_j\right),
\end{equation}
where the sum goes over all the proto-halo particles and $\delta_{ij}$ is the Kroenecker delta. The vector $\mathbf{x}$ corresponds to the particle position with respect to the centre-of-mass of the proto-halo. The eigenvalues of $I_{ij}$ are directly related to the axis $a\geq b\geq c$ of an ellipsoid with the same inertia tensor. This means it is possible to define the axis ratios $s=c/a$ and $q=c/b$ providing a measure of the proto-halo sphericity\footnote{In \citet{Lovell2014} only $s$ is used to distinguish between artefacts and real hales. We find it advantageous to use both $s$ and $q$ for the filtering procedure.}.

In Fig.~\ref{fig:halohistogram} we show histograms of haloes with ellipticity parameters $s$ (blue) and $q$ (red). The top row corresponds to the cases of CDM (left) and WDM with $m=0.25$ keV (right). While for CDM both parameters follow approximately Gaussian distributions, the WDM case exhibits secondary peaks at low values of $s$ and $q$. This is a clear indication for the presence of an artificial population of haloes with very elongated proto-haloes. The bottom row of Fig.~\ref{fig:halohistogram} illustrates the same statistics for the MDM models with mixing factor $f=0.8$ (left), $f=0.5$ (middle), and $f=0.2$ (right). The distributions do not show secondary peaks, but they are considerably broader and shifted towards lower values of $s$ and $q$ than in the case of CDM, indicating the presence of a subpopulation of artefacts in these models as well.

Based on the information of Fig.~\ref{fig:halohistogram}, we set cuts of $s_{\rm c}=0.24$ and $q_{\rm c}=0.34$ on the halo samples, removing all objects with smaller ellipticity parameters in all DM models. We will call this the {\it spericity cut} in the following.  In the WDM model, the values $s_{\rm c}$ and $q_{\rm c}$  lie at the minimum between the two peaks of real and artificial haloes, filtering out 60 percent of all haloes. In the case of CDM, on the other hand, the same values remove less than 5 percent of all haloes. This is a small loss considering the fact that changing the halo-finder can translate into ten percent differences in the abundance of haloes \citep{Knebe2011}.

The second step of the algorithm consists of additionally filtering out haloes with a poor match between simulations of the same initial field but different resolution. We again follow \citet{Lovell2014} and define the merit function
\begin{equation}\label{merit}
M = \frac{V_{\rm shared}}{\sqrt{V_{\rm LR}V_{\rm HR}}}
\end{equation}
for every proto-halo, where $V_{\rm LR}$ and $V_{\rm HR}$ are the proto-halo volumes in the low resolution (LR) and high resolution (HR) simulations, and where $V_{\rm shared}$ is the overlap of the two (i.e the shared volume). The merit function can be computed with any standard merger-tree algorithm based on particle IDs \citep[see][]{Srisawat2013}, as they can be used to link haloes between different simulations. Since standard merger-tree codes require the same particle numbers between analysed snapshots, we formally increase the particle number of the LR output by splitting every LR particle into $n$ particles corresponding to the HR particles of the same phase-space volume. This simple trick allows us to use the publicly available code {\tt MergerTree} \citep[out of the {\tt AHF}-package,][]{Knollmann2009} to calculate the number of shared particles between all haloes in the LR and HR simulations. This then trivially leads to the merit function of Eq.~\ref{merit} since every particle can be related to a small volume element in the initial conditions. 

\begin{figure}
\centering
\includegraphics[trim = 8mm 4mm 40mm 14mm, clip, scale=0.55]{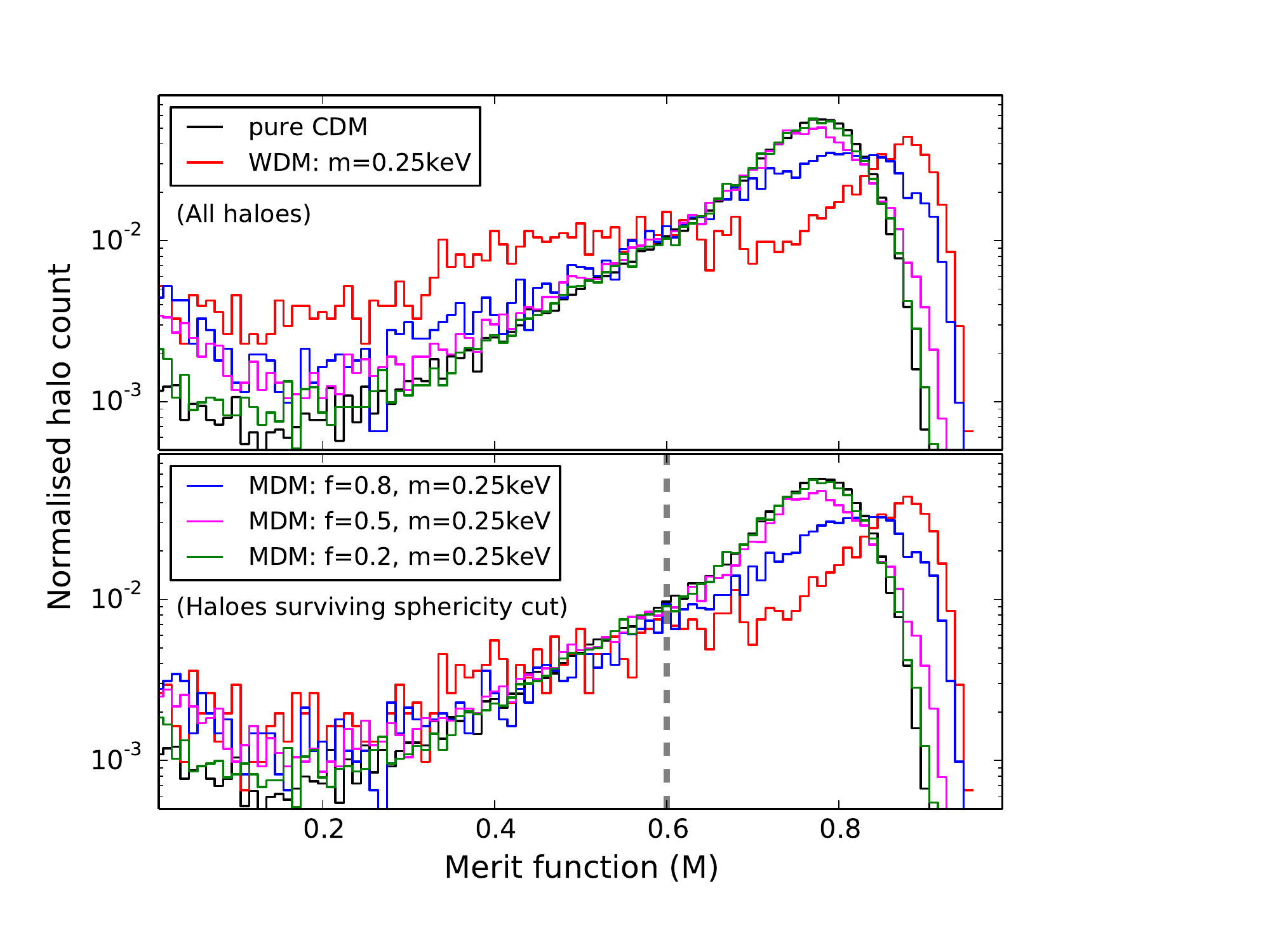}
\caption{\label{fig:halohistogram2}
Normalised halo histograms of the merit function (which measures the overlap of proto-haloes between simulations of different resolution). The top and bottom panels refer to the halo sample before and after the {\it sphericity cut}. The vertical dashed line illustrates the {\it resolution cut}.}
\end{figure}

In Fig.~\ref{fig:halohistogram2} we illustrate the histograms of haloes with respect to the merit function $M$ for CDM (black), WDM (red), and MDM (blue, green, and magenta). The top panel shows all haloes with more than 1000 particles, i.e before applying the {\it spehricity cut}. While the CDM distribution only exhibits one distinct peak at $M\sim0.8$, the WDM distribution has a peak at $M\sim0.9$ plus a secondary, much broader feature at $M<0.7$ which can be attributed to artefacts. The bottom panel of  Fig.~\ref{fig:halohistogram2} shows the same statistics for the haloes which are not filtered out by the {\it sphericity cut}. While the CDM distribution does hardly change, the broad feature at low $M$ in WDM is substantially smaller, indicating that most of the artefacts have already been filtered out. However, some artefacts survive the {\it sphericity cut} and still contribute to the histograms of the bottom panel. This is emphasised by the relative excess of WDM and MDM haloes at low values of $M$.

In order to filter out the remaining artefacts visible in the lower panel of Fig.~\ref{fig:halohistogram2}, we use the statistics of the merit function to apply a second selection cut of $M_{\rm c}=0.6$ (illustrated by the grey dashed line) which we will call the {\it resolution cut}. This further removal should capture most of remaining artefacts, while keeping the majority of the real haloes in the sample. For the cases of CDM and WDM the {\it resolution cut} filters out an additional 15 and 40 percent of the remaining haloes.

In summary, we note that applying solely the {\it sphericity cut} leads to rather conservative results in the sense that CDM haloes are hardly affected, but it does not filter out all of the WDM (and MDM) artefacts. Applying both the {\it sphericity} and the {\it resolution cut} filters out more artefacts, but it also removes some of the real objects. 

\begin{figure*}
\includegraphics[trim = 7mm 4mm 42mm 14mm, clip, scale=0.42]{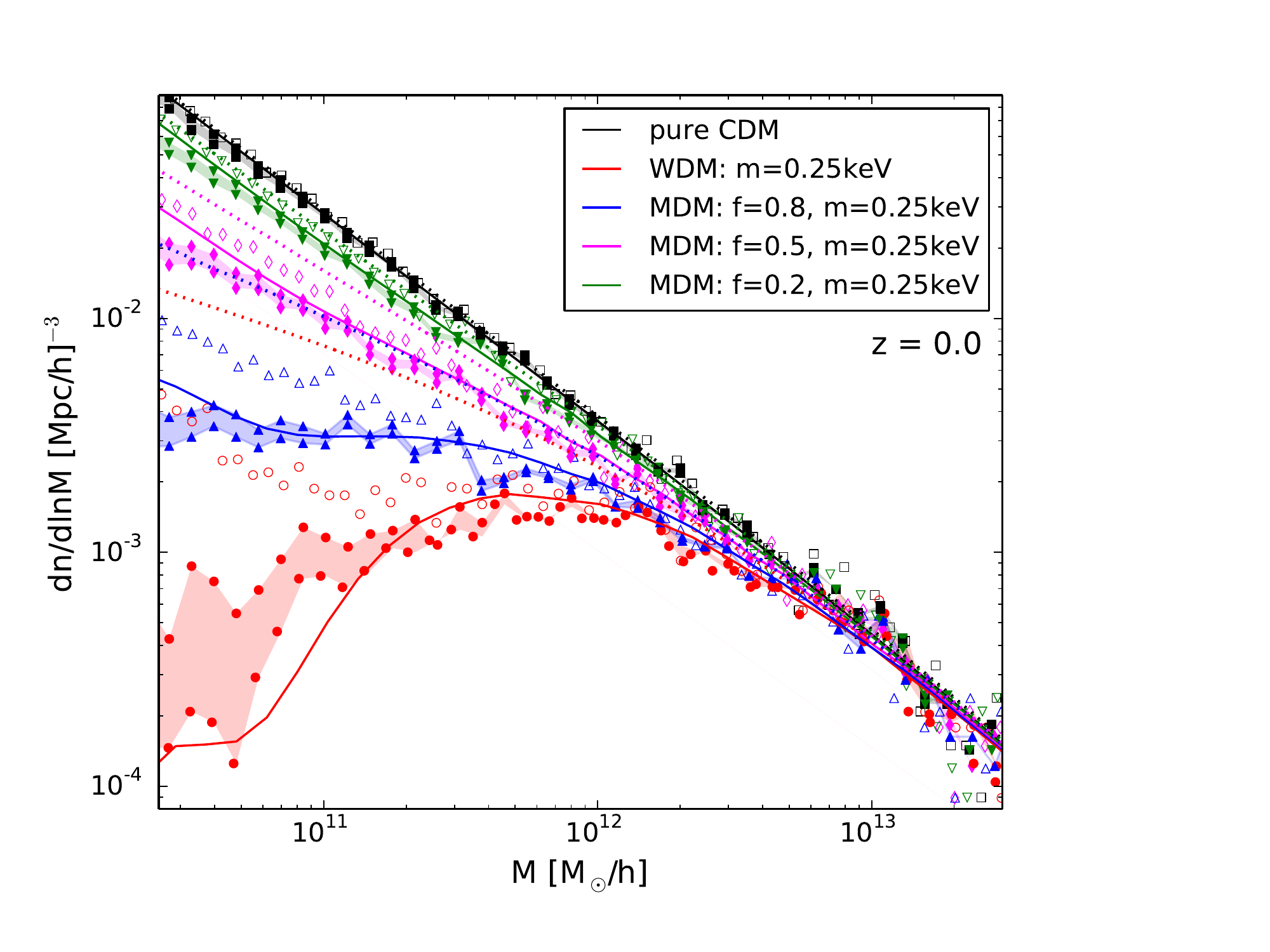}
\includegraphics[trim = 7mm 4mm 42mm 14mm, clip, scale=0.42]{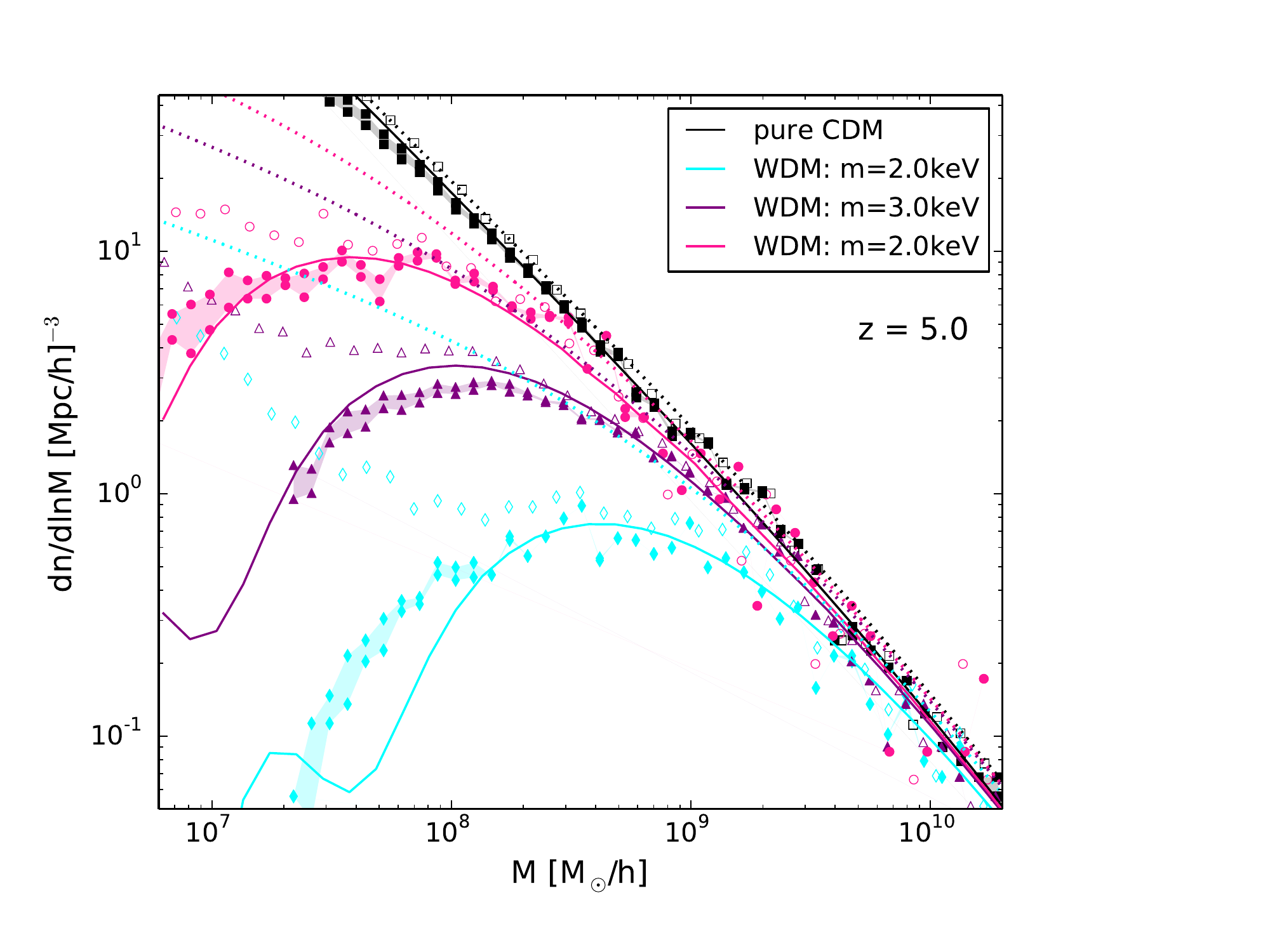}
\includegraphics[trim = 6mm 4mm 95mm 14mm, clip, scale=0.42]{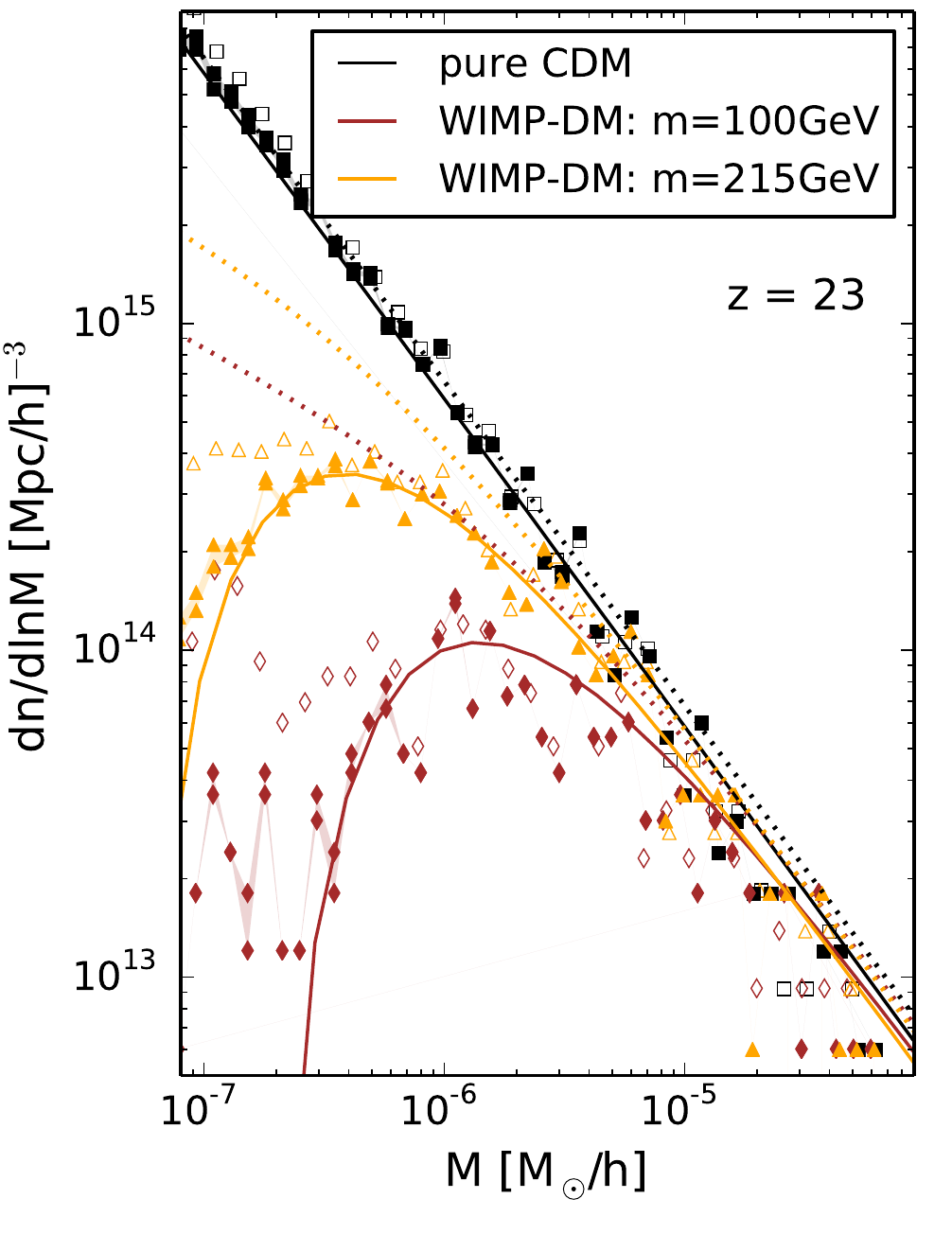}
\caption{Halo mass functions for various DM models. Left: CDM (black), WDM (red), and MDM (blue, magenta, green) at redshift zero. Middle: CDM (black) and WDM (cyan, purple, pink) at redshift 5. Right: pure CDM (black) and WIMP DM (brown, orange) at redshift 23. The colour-shaded regions account for the uncertainty due to the removal procedure of artefacts (see text). Solid lines represent the sharp-$k$ model, dotted lines the standard Sheth-Tormen mass function.
}\label{fig:massfct}
\end{figure*}

The halo mass function of Fig.~\ref{fig:massfct} illustrates how the method of removing artefacts affects the total halo abundance at different masses. Empty and filled symbols illustrate the mass function before and after the removal of haloes. Filled symbols appear twice illustrating the effect of only applying the {\it sphericity cut} (upper symbols) and both the {\it sphericity} and {\it resolution cut} (lower symbols). The coloured area in between therefore illustrates the inherent uncertainty of the method.

\section{Extended Press-Schechter model}\label{sec:EPS}
Many aspects of nonlinear structure formation can be analytically described by the extended Press-Schechter (EPS) approach, which assumes linear growth of perturbations followed by immediate halo formation above a certain threshold \citep[see][for a review]{Zentner2007}. The threshold is derived from an idealised spherical or ellipsoidal collapse calculation. Despite of these oversimplified assumption, the EPS model is able to reproduce fundamental statistics of structure formation, such as halo mass function, accretion history or bias \citep{Press1974,Bond1991,Sheth1999}. This is, however, only true for a pure CDM scenario where the initial power spectrum behaves as a quasi power-law which is never steeper than $k^{-3}$. As soon as the initial power spectrum is suppressed further (i.e. becomes steeper than $k^{-3}$), the standard EPS model completely fails to predict the right clustering. In this section, we show how the formalism needs to be adapted in order to give adequate predictions for cosmologies with arbitrary initial power spectra including steep cutoffs or gradual suppressions. We thereby follow and extend the method developed in \citet{Schneider2013}.

\subsection{Halo mass function: the standard approach}
The most used application and the starting point of the EPS model is the halo mass function. In the standard scenario it can be written as
\begin{equation}\label{sEPSmassfct}
\frac{dn}{d\ln M}= \frac{1}{2}\frac{\bar{\rho}}{M}f(\nu)\frac{d\ln\nu}{d\ln M},
\end{equation} 
where $n$ is the number density of haloes, $M$ the halo mass, $\nu$ the peak-height of perturbations, and $\bar\rho$ the average density of the universe \citep{Press1974}. The first crossing distribution $f(\nu)$ is obtained with the excursion-set approach, which follows random walk trajectories counting events of first up-crossing of the collapse threshold \citep{Bond1991}. The spherical collapse model predicts a constant threshold, leading to
\begin{equation}\label{PSdist}
f(\nu) = \sqrt{\frac{2\nu}{\pi}}e^{-\nu/2}.
\end{equation}
The more realistic case of ellipsoidal collapse, on the other hand, leads to a mass dependent threshold and a first-crossing distribution of the form
\begin{equation}\label{STfirstcrossing}
f(\nu) = A \sqrt{\frac{2q\nu}{\pi}}\left[1+(q\nu)^{-p}\right]e^{-q\nu/2}
\end{equation}
with $A=0.3222$, $p=0.3$, and $q=1$ \citep{Sheth1999}. The peak height $\nu$ is defined as
\begin{equation}
\nu=\frac{\delta_{c,0}^2}{S_\chi(R)D^2(a)},
\end{equation}
where $\delta_{c,0}=1.686$ and where
\begin{equation}
D(a)=\frac{5\Omega_m}{2}H(a)\int\frac{da}{a^3H(a)^3} 
\end{equation}
is the universal growth factor of perturbations. Here we have introduced the Hubble parameter $H=\dot{a}/a$. Finally, the variance $S_\chi$ is defined as
\begin{equation}\label{variance}
S_\chi(R)=\int \frac{d^3\mathbf{k}}{(2\pi)^3}P_\chi(k)W^2(k|R),
\end{equation}
where $\chi$ stands for the DM scenario imposed by the linear power spectrum $P_\chi(k)$. There is a filter function in Fourier-space $W(k|R)$ appearing in Eq.~\ref{variance}, which is a priori unconstrained. One obvious choice used in the standard EPS approach is a top-hat function in real space with enclosed mass $M=4\pi\bar{\rho}R^3/3$, transforming into
\begin{equation}\label{tophatfilter}
W(k|R)=\frac{3}{(kR)^3}\left[\sin(kR)-3\cos(kR)\right]
\end{equation}
in Fourier space. Other common choices for $W(k|R)$ are a sharp-$k$ filter (a tophat function in Fourier space) or a Gaussian filter \citep{Bond1991}.

The usual recipe to construct the EPS halo mass function consists of combining Eq. (\ref{sEPSmassfct} - \ref{tophatfilter}) leading to very good agreement with CDM simulations for halo masses below $M=10^{13}$ $h^{-1}\rm M_{\odot}$.  Significantly larger haloes, on the other hand,  are under-predicted by the model. Motivated by this discrepancy at the largest scales, \citet{Sheth1999} heuristically modified the first-crossing distribution by setting $q=0.707$, what leads to very good agreement with simulations over all mass scales, but it technically consists of allowing for one free parameter. This modified model is commonly referred to as the Sheth-Tormen mass function.

In Fig. \ref{fig:massfct} we show the expected good agreement between the Sheth-Tormen mass function (dotted lines) and simulations of a CDM cosmology (black squares), which holds over a wide range of halo masses and redshifts. However, Fig. \ref{fig:massfct} also shows the complete failure of the Sheth-Tormen mass function to reproduce the halo abundance of WDM, MDM or WIMP DM simulations (coloured symbols) around the scale of suppression (the coloured area between the two sets of filled symbols designate the uncertainty in the method to filter out artificial haloes, the empty symbols illustrate the raw data, including real haloes and artefacts; see Sec.~\ref{sec:remart} for more details).

\subsection{Halo mass function: the sharp-$k$ model}
The inability of the standard EPS approach to model cosmologies with suppressed power spectra has been reported by several papers in the past \citep{Barkana2001,Benson2013,Schneider2013,Hahn2014}, and different methods to solve this problem have been proposed. A simple and elegant way of modifying the model is to change the filter from a top-hat function in real space to a top-hat function in Fourier space, called sharp-$k$ filter, i.e.
\begin{equation}\label{sharpkfilter}
W(k|R)=\Theta(1-kR),
\end{equation}
where $\Theta$ is the Heaviside step function. This possibility was initially discussed in a paper by \citet{Bertschinger2006} and has been adapted to WDM by \citet{Benson2013} and \citet{Schneider2013}. The reason why the sharp-$k$ filter does a better job than the standard top-hat filter lies in the asymptotic behaviour of Eq. (\ref{variance}) towards small radii. With a sharp-$k$ filter, this integral naturally depends on the shape of the power spectrum for any radius. With a top-hat filter, on the other hand, the integral becomes completely insensitive to the shape of the power spectrum as soon as the latter decreases faster than $k^{-3}$. Instead, the mass function becomes solely driven by the shape of the filter function and strongly deviates from the halo abundance in simulations \citep[see][for a more detailed discussion]{Schneider2013}.

For the case of a sharp-$k$ filter, the functional form of the halo mass function (i.e. Eq.~\ref{sEPSmassfct}) can be further simplified. A straight-forward calculation leads to the expression
\begin{equation}
\frac{dn_{\rm SK}}{d\ln M}=\frac{1}{12\pi^2}\frac{\bar\rho}{M}\nu f(\nu)\frac{P(1/R)}{\delta_{\rm c}^2R^3}.
\end{equation}
For the sake of simplicity, we have dropped the $\chi$-index standing for the DM model. Together with Eqs. (\ref{STfirstcrossing}), (\ref{variance}), and (\ref{sharpkfilter}) as well as with an appropriate relation between filter scale and mass, we get a closed set of equations yielding the halo number density per mass scale.

The sharp-$k$ filter has an important drawback we haven't discussed so far, namely that it has no well defined mass $M$ associated to its filter scale $R$. This means that, apart from the proportionality $M\propto R^3$ due to the spherical symmetry of the filter, the halo mass is unconstrained. Introducing a free parameter $c$, the relation between filter scale and mass can be written as
\begin{equation}\label{MvsR}
M=\frac{4\pi}{3}{\bar\rho}(cR)^3,
\end{equation} 
where $c=2.5$ gives the best match to simulations\footnote{The best value for c depends somewhat on the halo-definition and the halo-finding routine. In Schneider et al 2013 we used a slightly higher value $c=2.7$.}. The calibration of the parameter $c$ is done once and does not change for different DM scenarios.

The sharp-$k$ model has two main advantages compared to the standard EPS approach: First, it consistently uses the same filter (the sharp-$k$ function) for the entire excursion-set calculation, while the EPS approach usually relies on the sharp-$k$ filter to calculate the first crossing distribution and on the tophat filter to connect it to the number density of haloes. Second, the sharp-$k$ method is based on the original first-crossing distribution from ellipsoidal collapse, while the Sheth-Tormen model introduces a heuristic parameter adjustment. The second advantage is counterbalanced by the fact that the sharp-$k$ filter does not have a well defined mass. At the end, both the sharp-$k$ and the Sheth-Tormen model have one free parameter, which needs to be adjusted to simulations.

The halo mass function from the simulations (symbols), the Sheth-Tormen model (dashed lines), and the sharp-$k$ model (solid lines) are all plotted in Fig. \ref{fig:massfct}. For the case of CDM (black), the sharp-$k$ mass function closely follows both the simulation measurements and the Sheth-Tormen model. For the case of WDM (red, cyan, purple, pink), MDM (green, magenta, blue), and WIMP DM (brown, orange), the sharp-$k$ mass function gives a reasonably good match to simulations, while the Sheth-Tormen approach fails to match the flattening or the turnaround visible in simulations.

In  \citet{Schneider2013} the sharp-$k$ model has been reported to underestimate the halo abundance when the suppression scale lies in the exponential tail of the halo mass function (i.e. for $\nu\gg1$), which generally happens at very high redshift. It turns out, however, that this discrepancy between the sharp-$k$ model and the data is greatly reduced for haloes defined by a spherical overdensity instead of a friends-of-friends linking criterion \citep[see][for a comparison of the two]{Watson2013}. We therefore do not use the correction model proposed by \citet{Schneider2013}.

\subsection{Conditional mass function}
Another important application of the EPS model is the conditional mass function, which gives the abundance of haloes per mass and look-back redshift $z_1$, eventually ending up in a single host halo at redshift $z_0$. As the conditional mass function provides a connection between haloes at different redshifts, it acts as the starting point of more evolved quantities such as the halo collapse redshift, the number of satellites, and halo merger trees. The conditional mass function is given by
\begin{equation}\label{condMF}
\frac{dN(M|M_0)}{d\ln M}=-\frac{M_0}{M} Sf(\delta_{\rm c},S | \delta_{\rm c,0},S_0)\frac{d\ln S}{d\ln M}
\end{equation}
\citep{Lacey1993}. For the sharp-$k$ model this can be simplified to
\begin{equation}
\frac{dN_{\rm SK}(M|M_0)}{d\ln M}=\frac{1}{6\pi^2}\frac{M_0}{M} f(\delta_{\rm c},S | \delta_{\rm c,0},S_0) \frac{P(1/R)}{R^3},
\end{equation}
where the filter scale $R$ and the mass $M$ are related by Eq. (\ref{MvsR}).

The conditional first-crossing distribution again depends on the assumed model for nonlinear collapse. The case of spherical collapse is given by
\begin{equation} \label{condPSdist}
f(\delta_{\rm c},S | \delta_{\rm c,0},S_0)=\frac{\left(\delta_c-\delta_{c,0}\right)}{\sqrt{2\pi(S-S_0)}}\exp\left[-\frac{(\delta_c-\delta_{c,0})^2}{2(S-S_0)}\right],
\end{equation}
which is a simple recalibration of Eq. (\ref{PSdist}), obtained by shifting the starting point of trajectories from $(0,0)$ to $(\delta_{c,0},S_0)$ in the $(\delta_c-S)$ plane. In the case of ellipsoidal collapse, the threshold depends on the variance $S$ and no simple recalibration is possible. As a consequence, every point in the $(\delta_c-S)$ plane requires an independent excursion-set calculation, and no general analytical expression exists for the conditional first-crossing distribution \citep{Sheth1999}. To keep it simple, we therefore use the spherical collapse model (i.e. Eq.~\ref{condPSdist}) whenever we compute the conditional mass function in this paper. 

\begin{figure}
\includegraphics[trim = 8mm 4mm 40mm 14mm, clip, scale=0.55]{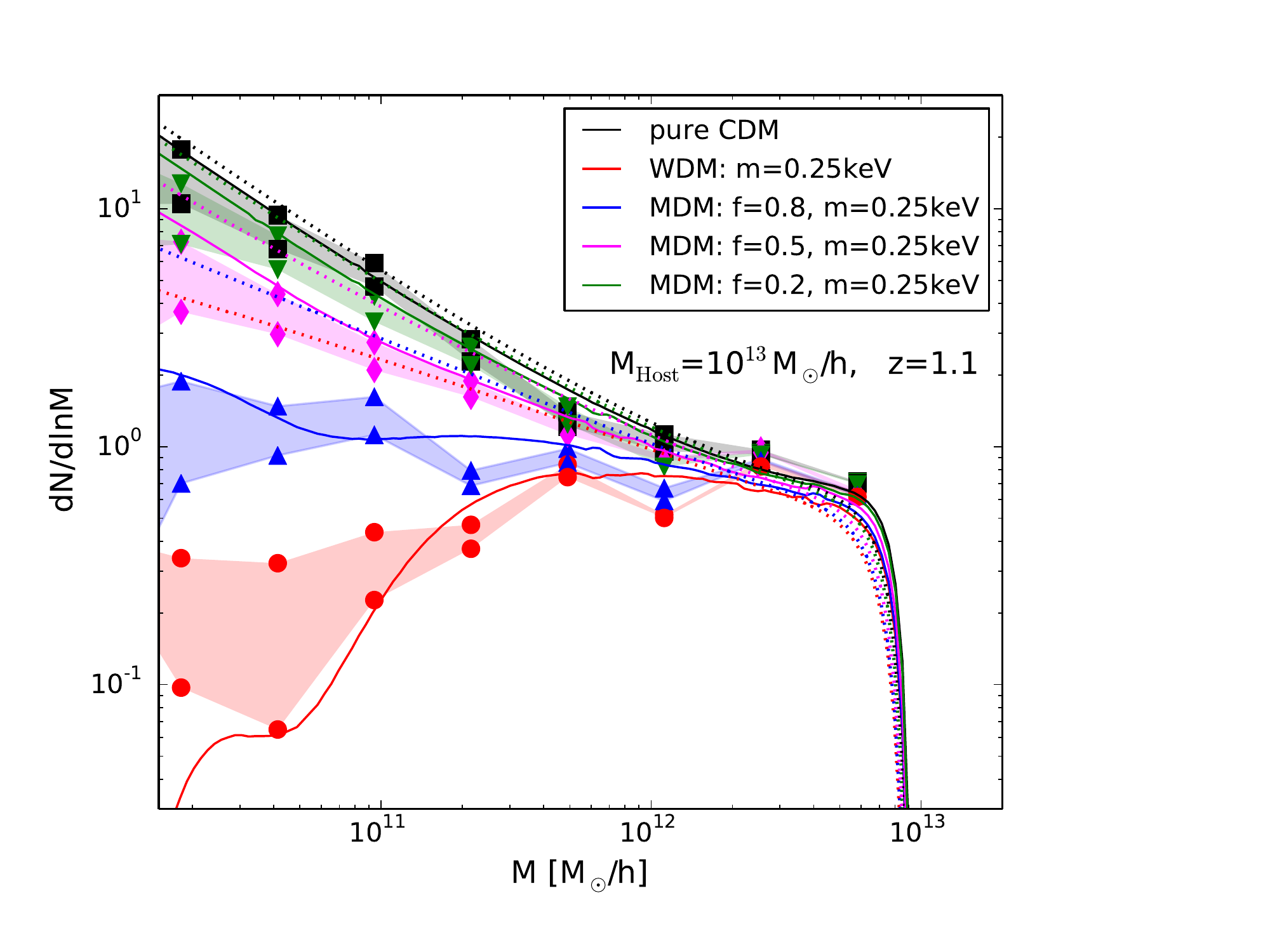}
\caption{Conditional mass functions for a $M_0=10^{13}$ $h^{-1}\rm M_{\odot}$ host and a look-back redshift of $z=1.1$. Coloured symbols refer to simulation outputs (with circumjacent shaded regions representing the uncertainty due to artefact subtraction), while the solid and dotted lines represent the sharp-$k$ model and the standard Press-Schechter model, respectively. The colour-coding is the same as in the previous plots.}\label{fig:condMF}
\end{figure}

In Fig. \ref{fig:condMF} the conditional mass function is plotted for the case of a $10^{13}$ $h^{-1}\rm M_{\odot}$ host and a look-back redshift of $z=1.1$. The symbols correspond to the average number of progenitor haloes out of 84 host systems with masses close to $10^{13}$ $h^{-1}\rm M_{\odot}$ (upper symbols stand for the conservative and lower symbols for the more radical method to filter out artefacts, see Sec.~\ref{sec:remart} for further information). The solid and dotted lines are predictions from the sharp-$k$ and the standard EPS approach, respectively. While the sharp-$k$ model gives a good match to simulations of all DM scenarios, the standard EPS model works for CDM but fails for all scenarios with suppressed small-scale perturbations.

\subsection{Estimating the number of dwarf satellites}
Each DM scenario has to produce a sufficient amount of substructures to account for the observed Milky Way satellites. While some (or most) of the substructures could be dark due to inefficient star formation, fewer substructures than observed means the failure of the DM scenario. In the case of WDM, comparing numbers of simulated substructures with observed satellites has lead to tight constraints on the thermal particle mass ruling out masses below 2 keV \citep{Polisensky2011,Kennedy2014}.

The EPS approach can be used to estimate the average number of dwarf galaxies orbiting a galaxy like the Milky Way (MW). This means it is possible to check whether a certain DM scenario is likely to be in agreement with observations without running expensive numerical zoom-simulations of a MW halo. In principle, finding the number of progenitors from an EPS approach consists of constructing the full merger-tree of the host-halo and counting all branches directly connected to the trunk. This is a rather cumbersome calculation which lies beyond the scope of this paper. Instead, we follow a simplified procedure presented in \citet{Giocoli2008} and adapt it to the sharp-$k$ model.

The total number of progenitors over all redshifts $N_{\rm Sat}(\delta_c>\delta_{c,0})$ can be computed by integrating the conditional mass function over $\delta_c$, i.e.
\begin{equation}\label{Nprog}
\frac{dN_{\rm Sat}}{d\ln M}=\frac{1}{N_{\rm norm}}\int_{\delta_{c,0}}^{\infty}\frac{dN_{\rm SK}}{d\ln M}d\delta_c
\end{equation}
with $N_{\rm norm}=1$ for now. This integral over-counts the actual number of progenitors (and therefore the expected number of substructures) because it includes multiple counts of structures simultaneously existing at different redshifts. This issue can be accounted for by normalising the integral so that Eq. (\ref{Nprog}) matches the outcome of $\Lambda$CDM simulations. We use data based on the Aquarius project of a simulated MW halo with mass $M_0=1.77\cdot 10^{12}$ $h^{-1}\rm M_{\odot}$, which has 157 satellites  of $M>10^8$ $h^{-1}\rm M_{\odot}$ \citep{Lovell2014}. Integrating Eq. (\ref{Nprog}) over the mass range between $10^8$ $h^{-1}\rm M_{\odot}$ and $M_0$ and equating it to the number of simulated satellites, leads to the normalisation constant $N_{\rm norm}=44.5$. For the case of a sharp-k filter Eq. (\ref{Nprog}) then becomes
\begin{equation}\label{numsat}
\frac{dN_{\rm Sat}}{d\ln M}=\frac{1}{44.5}\frac{1}{6\pi^2}\left(\frac{M_0}{M}\right)\frac{P(1/R)}{R^3\sqrt{2\pi(S-S_0)}}.
\end{equation}
This relation provides an estimate for the number of satellites per mass $M$ of a host halo with mass $M_0$, where $S$ and $S_0$ are the corresponding variances and $P(1/R)$ is the linear power spectrum at the inverse of the filter scale $R$. A further integration over $M$ finally leads to the total amount of satellites for a DM scenario with arbitrary power spectrum.

The model is clearly an over-simplification and cannot capture the full complexity of hierarchical structure formation. First, it does not give the final subhalo mass prior to the merger with the host, but rather some average mass of the subhalo formation history. Second, it completely ignores tidal striping which reduces the mass of substructures and becomes significant as soon as a satellite passes close to the centre of the host.

\begin{table}
\ra{1.2}
   \centering{
\begin{tabular}{llcc}
  \hline
  Scenario & & $N_{\rm Sat}$ & $N_{\rm Sat}$  \\
  & & (EPS model) & (Simulation)\\
  \hline
  WDM: & m=1.5 keV & 9 & 12 \\
  WDM: & m=2.0 keV & 25 & 27 \\
  WDM: & m=2.3 keV & 32 & 30 \\
  CDM:  & pure & 158 & 158 \\
  \hline
\end{tabular}}
\caption{Number of predicted Milky-Way satellites with mass above $10^8$ $h^{-1}\rm M_{\odot}$ from the model (Eq.~\ref{numsat}) and from simulations by \citet{Lovell2014}. For the case of CDM the perfect agreement is by construction, since the model is calibrated to this number.}\label{tab:Nsat}
\end{table}

In Table \ref{tab:Nsat} we give the number of satellites with $M>10^8$ $h^{-1}\rm M_{\odot}$ predicted by the EPS model for different WDM scenarios. The numbers agree surprisingly well with satellite counts from WDM simulations of \citet{Lovell2014}. The predicted number of satellites lies close to the outcome of simulations well within the expected statistical fluctuation between different host haloes. These results strongly suggests that despite its over-simplifications, the model (i.e Eq.~\ref{numsat}) can be used to estimate the amount of dwarf satellites for scenarios with suppressed power spectra. It consists of a very simple and useful test to check if a DM scenario is likely to be ruled out by dwarf galaxy counts, or if it has the potential to account for potential {\it small scale problems} of $\Lambda$CDM.

As an example, we can use the model developed above to constrain the mixed DM (MDM) parameter space. There are 11 classical satellites plus additional 15 satellites recently observed by the Sloan Digital Sky Survey (SDSS) within the viral radius of the MW \citep{Wolf2010}. As SDSS covers only 28\% of the sky, the latter number needs to by multiplied by 3.5 to account for full sky coverage (assuming spherically distributed dwarfs). Altogether this leads to an estimate of $11+50=61$ dwarf satellites from observations \citep[see][for more details about how to estimate the number of MW satellites]{Polisensky2011}.

The total mass of these objects are very difficult to infer, since the only information about the dark matter halo comes from stellar kinematics at around the half-light radius. Fitting different possible profiles and extrapolating to the viral radius, \citet{Brooks2014} find that the masses of all classical and SDSS dwarfs should lie above $10^8$ $h^{-1}\rm M_{\odot}$. We therefore use $M_{\rm min}=10^8$ $h^{-1}\rm M_{\odot}$ as a lower limit for our analysis.

The predicted number of satellites is very sensitive to the host mass, which is uncertain by a factor of a few in the case of the MW. In our analysis we take the bounds $5.5\times10^{11}<M_{0}<3.2\times10^{12}$ $h^{-1}\rm M_{\odot}$ on the MW mass obtained by \citet{Guo2010}.

\begin{figure}
\includegraphics[trim = 8mm 30mm 39mm 14mm, clip, scale=0.55]{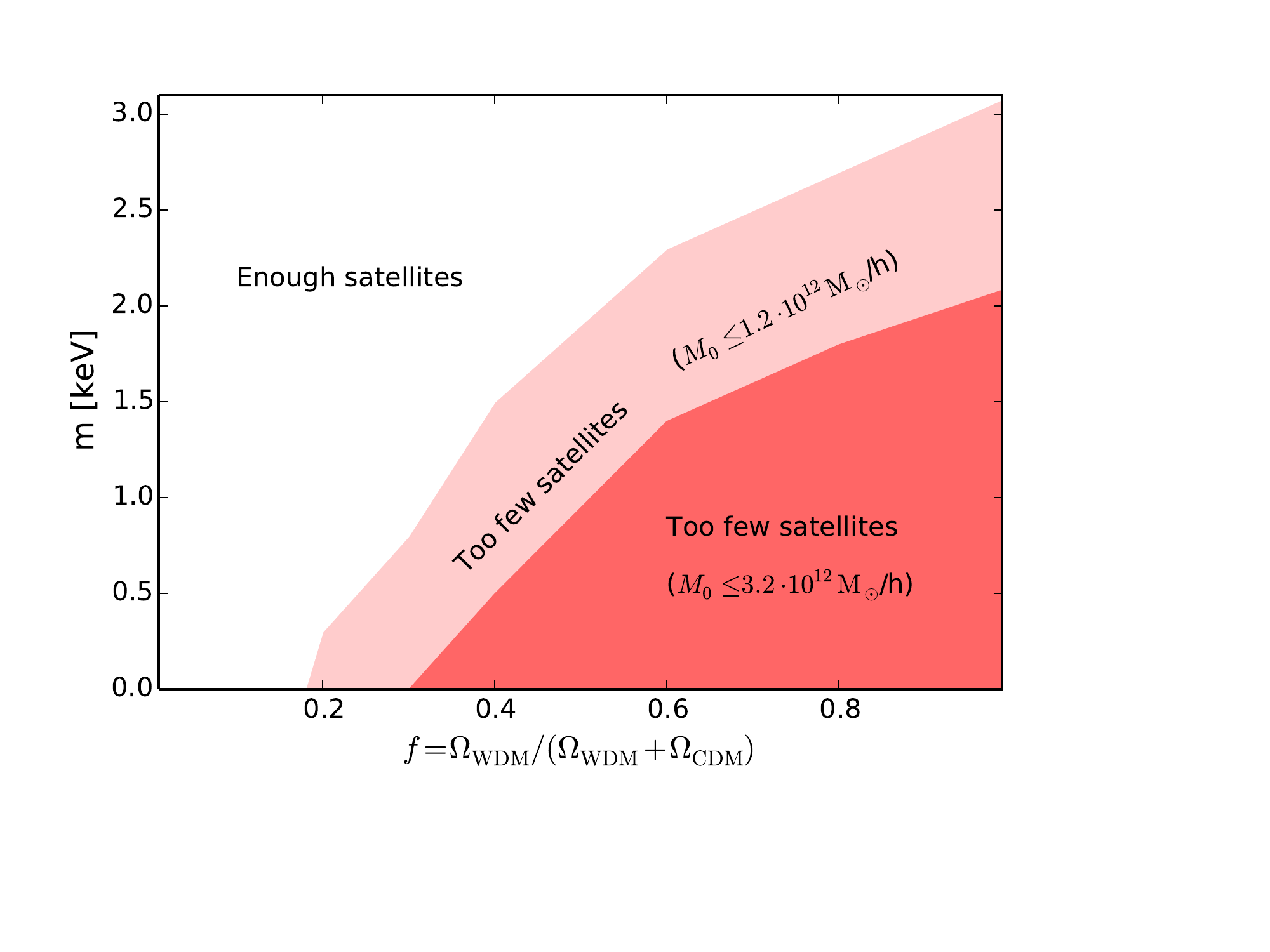}
\caption{Exclusion plot of mixed DM (MDM) based on satellite counts, assuming a Milky-Way mass of $1.2\times10^{12}$ and $3.2\times10^{12}$ $h^{-1}\rm M_{\odot}$ \citep[average and maximum estimates for the Milky-Way mass from][]{Guo2010}.}\label{MDMconstraints}
\end{figure}

Fig.~\ref{MDMconstraints} provides constraints on the mixed DM scenario obtained by combining the estimated satellite count with predictions from Eq.~(\ref{numsat}). The constraints depend on both the thermal mass ($m$) of the warm/hot DM part and the mixing fraction ($f$). As smaller host-haloes have fewer satellites, lower estimates of the MW mass translate into stronger constraints. We account for this by investigating exclusion limits for two MW masses, one corresponding to the mean value ($M_{0}\leq1.2\times10^{12}$ $h^{-1}\rm M_{\odot}$, pink shaded region) and one to the maximum value ($M_{0}\leq3.2\times10^{12}$ $h^{-1}\rm M_{\odot}$, red shaded region) given by \citet{Guo2010}.

For the extreme case of $f=1$ (corresponding to WDM), the constraints lie at 2-3 keV, which is roughly in agreement with estimates from \citet{Polisensky2011} and \citet{Kennedy2014}. On the other hand, MDM with fractions smaller than $f\sim 0.2-0.3$ cannot be constrained by satellite counts. A similar lower limit on $f$ was found by \citep{Anderhalden2013} based on satellite counts in MDM simulations and by \citet{Boyarsky2009} based on data from the Lyman-$\alpha$ forest.


\section{Concentration-mass relation}\label{sec:concentrations}
The concentration of a halo is defined as the ratio between virial radius and scale radius, assuming a broken power-law profile, where the virial radius delimitates the system and the scale radius defines the break between the two power-law regimes. Usually an NFW profile  \citep{Navarro1997} is assumed, which has an inner slope of $r^{-1}$ and an outer slope of $r^{-3}$. In a pure CDM scenario, halo concentrations are slightly mass dependent, decreasing on average towards larger halo masses with an important scatter between individual haloes. In WDM scenarios, halo concentrations have been reported to increase, turn over, and decrease on average towards large masses \citep{Bode2001,Avila-Reese2001,Schneider2012}.

The concentration-mass relation is a direct consequence of the connection between concentrations and the halo accretion history, as shown explicitly in numerical studies of pure CDM cosmologies \citep{Wechsler2002,Ludlow2014}. In a Universe governed by hierarchical clustering, small haloes collapse first while the average density is high what translates into an increased value for the concentration. This connection has been utilised in various analytically motivated models for the concentration-mass relation \citep{Navarro1997,Bullock2001,Eke2001}.

Arguably the most widely used model for the concentration-mass relation has been developed by \citet{Bullock2001} and assumes a concentration of
\begin{equation}\label{DcollapseBul}
c(z)=K\frac{(1+z)}{(1+z_c)},\hspace{0.5cm}\langle D(z_c)\rangle_\chi=\frac{\delta_{\rm c,0}^2}{S_{\chi}(FM)},
\end{equation}
with the free parameters K and F. The collapse redshift $z_c$ is derived from the second relation, where  $\langle D(z_c)\rangle_{\chi}$ stands for the growth factor of collapse of an average perturbation, assuming a DM scenario $\chi$. The Bullock-model approximately works for pure CDM, but it breaks down for the case of WDM where it predicts the concentration-mass relation to become constant instead of the observed downturn towards small masses \citep{Eke2001,Schneider2012}. 

The reason for this failure comes from an inaccuracy in the estimation of the redshift of collapse. While the Bullock model uses the average collapse redshift of perturbations with mass $M$ (i.e the second relation of Eq. \ref{DcollapseBul}), what is actually required is the average collapse redshift of a halo that has collapsed beforehand and still exists today. This subtle difference of perspective becomes important for large masses ($FM>M_*$, with $M_*$ being defined by the relation $S(M_*)=\delta_{c,0}^2$) where the Bullock model assigns negative collapse redshifts which is inherently contradictory. The inaccuracy of the Bullock model becomes more important for WDM, where haloes around the suppression scale collapse out of very shallow perturbations (that have been nearly entirely destroyed by free-streaming) and, therefore, only exist for a short time before they get accreted by a larger halo. As a result, small haloes which survive until today tend to have very low collapse redshifts. In the following, we derive a new relation for the average halo collapse redshift which takes this effect into account.

\subsection{Halo collapse redshift}
Halo collapse is not an instantaneous event but a lengthy processes of accretion and merging, and defining a formation time is therefore intrinsically ambiguous. The usual definition for the collapse redshift $z_c$ is the moment when a halo has accreted a fraction $F$ of its final mass $M$. Every halo has an individual accretion history and a different collapse redshift.

The distribution of collapse redshifts ($g_{\chi}$) for all progenitors of one halo with mass $M$ is simply given by the conditional first-crossing distribution
\begin{equation}\label{probfct}
g_{\chi}(\delta_c)=f\left[\delta_c,S_\chi(FM) | \delta_{c,0},S_\chi(M)\right],
\end{equation}
where $\delta_c=\delta_{c,0}/D(z_c)$ is the redshift dependent collapse threshold of a given progenitor (with $\delta_{c,0}=1.686$ in the spherical collapse model). Eq. (\ref{probfct}) is a well behaved probability function with an average value 
\begin{equation}\label{pertdist}
\langle \delta_c \rangle = \frac{\int d\delta_c\delta_c g_\chi(\delta_c)}{\int d\delta_c g_\chi(\delta_c)}.
\end{equation}
Solving this equation explicitly with the help of Eq. (\ref{pertdist}) yields a relation for the average growth-factor
\begin{equation}\label{Dcollapse}
\langle D(z_c)\rangle_{\chi}=\left[ 1+\sqrt{\frac{\pi}{2}}\frac{1}{\delta_{c,0}}\sqrt{S_\chi(FM)-S_\chi(M)}\right]^{-1}
\end{equation}
(for a given DM scenario $\chi$) from which it is straight forward to derive the corresponding redshift of collapse. Eq.~(\ref{Dcollapse}) is a clear improvement compared to Eq.~(\ref{DcollapseBul}) of the Bullock model, since it provides a reasonable value for the collapse redshift even in the case $FM>M_*$. Additionally, it allows the $z_c$-M relation (and therefore the concentration-mass relation as we will see in the next section) to turn over towards small masses, provided the function $S_\chi(M)$ becomes flat enough which is the case for WDM and some MDM scenarios. 

\begin{figure}
\includegraphics[trim = 8mm 4mm 40mm 14mm, clip, scale=0.55]{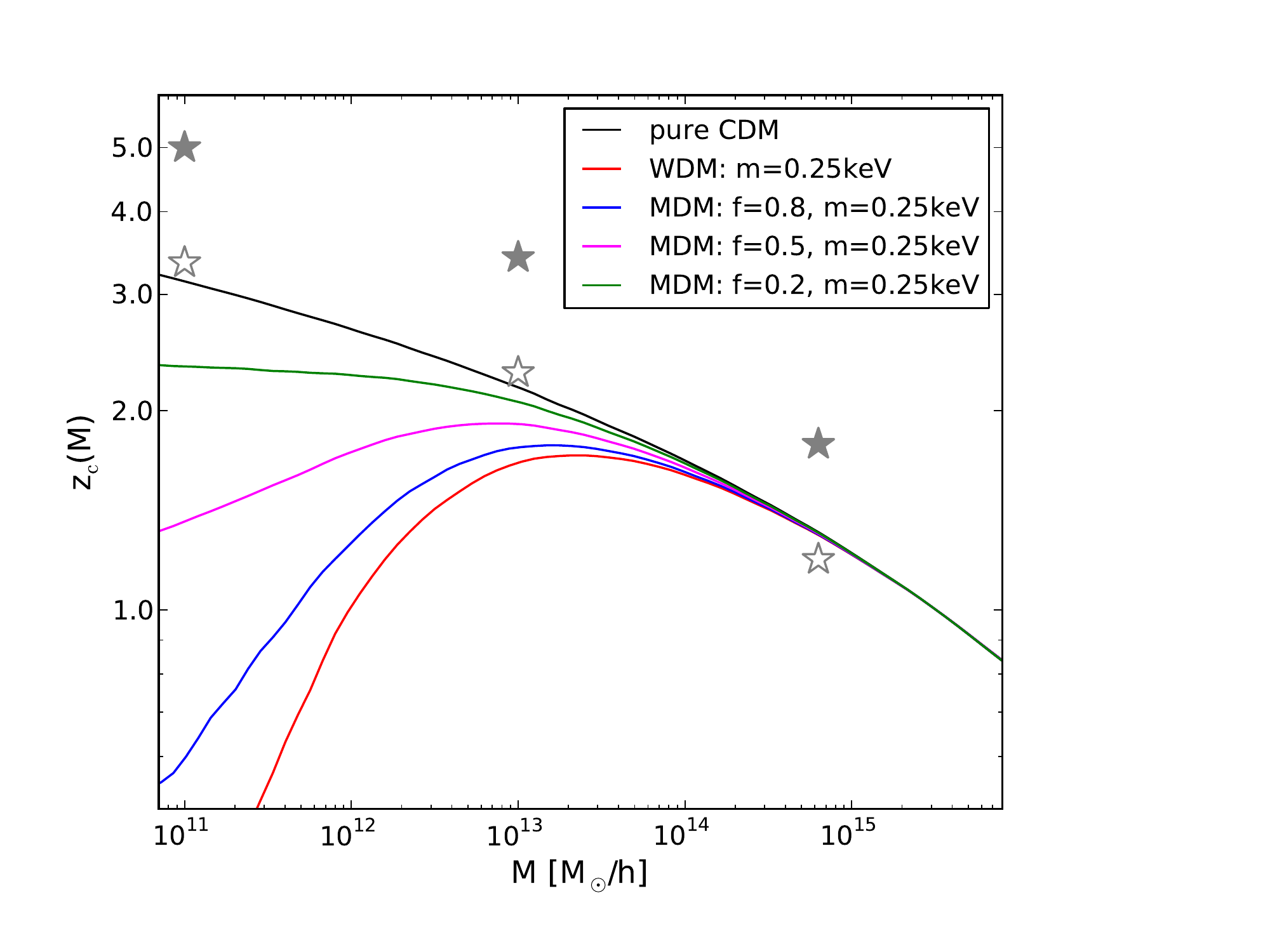}
\caption{Redshift of collapse of haloes with mass $M$ at redshift zero. The curves are drawn from Eq. (\ref{Dcollapse}) with $F=0.05$. The grey stars are extracted from measured halo accretion rates in CDM simulations \citep{Zhao2009}. Filled stars correspond to the absolute values, empty stars are normalised to the black curve.}\label{fig:z_collapse}
\end{figure}

The problem with the derivation above is that does not yield the collapse redshift of the main halo but rather an average collapse redshift of all its progenitors. In general, the true halo collapse redshift can only be determined by constructing the full EPS halo merger-tree, which lies beyond the scope of this work\footnote{Assuming the special case of $F\geq0.5$, it is possible to compute the collapse distribution (and therefore the average collapse redshift) by integrating the conditional mass function (\ref{condMF}) over the mass range $[FM,M]$ followed by a derivative with respect to $\delta$. This is, however, not a valuable option for the purpose of fitting concentrations, since $F$ needs to act as a free parameter in order to reproduce the measurements from simulations.}.

\begin{figure*}
\includegraphics[trim = 12mm 4mm 98mm 14mm, clip, scale=0.44]{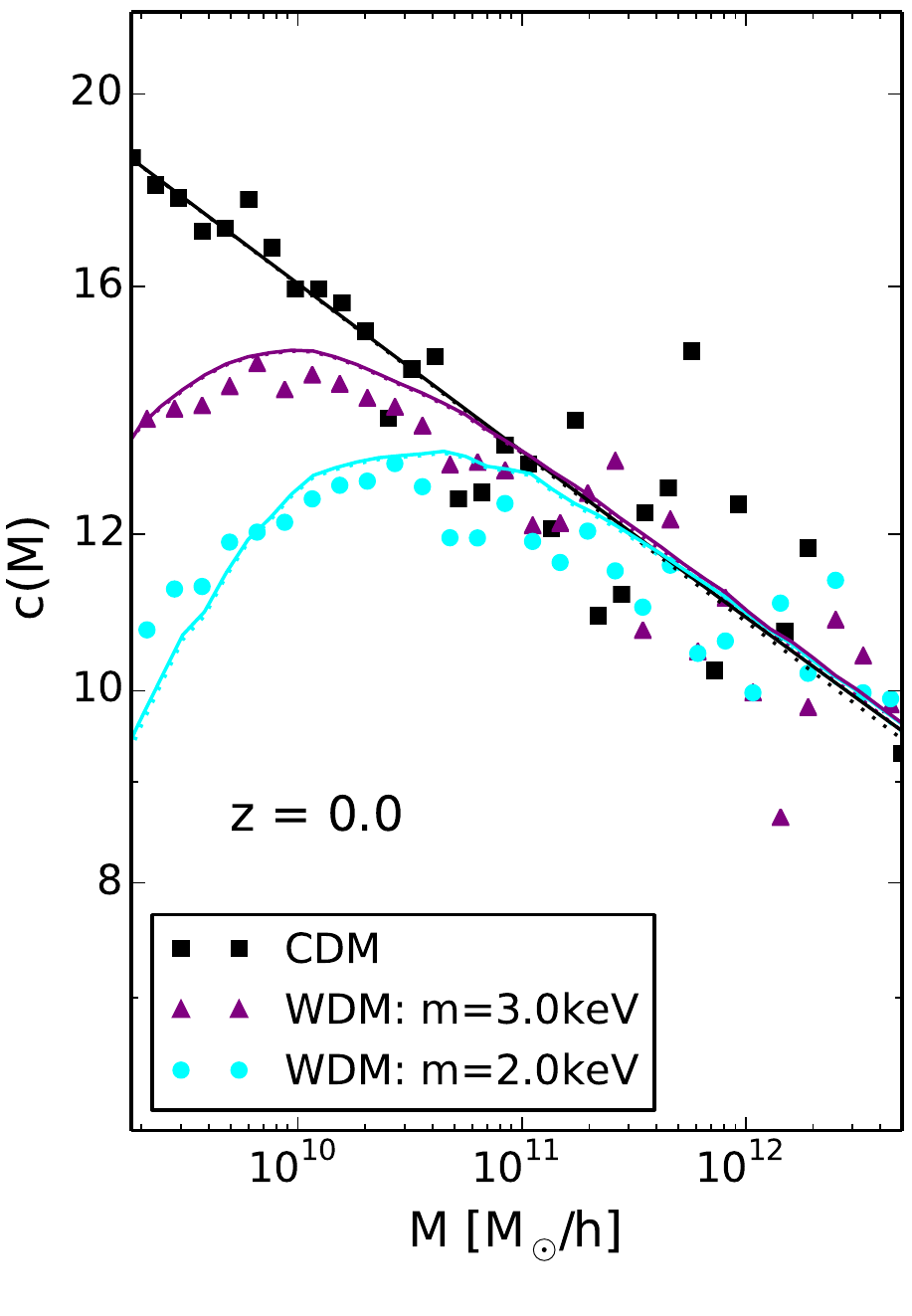}
\includegraphics[trim = 12mm 4mm 42mm 14mm, clip, scale=0.44]{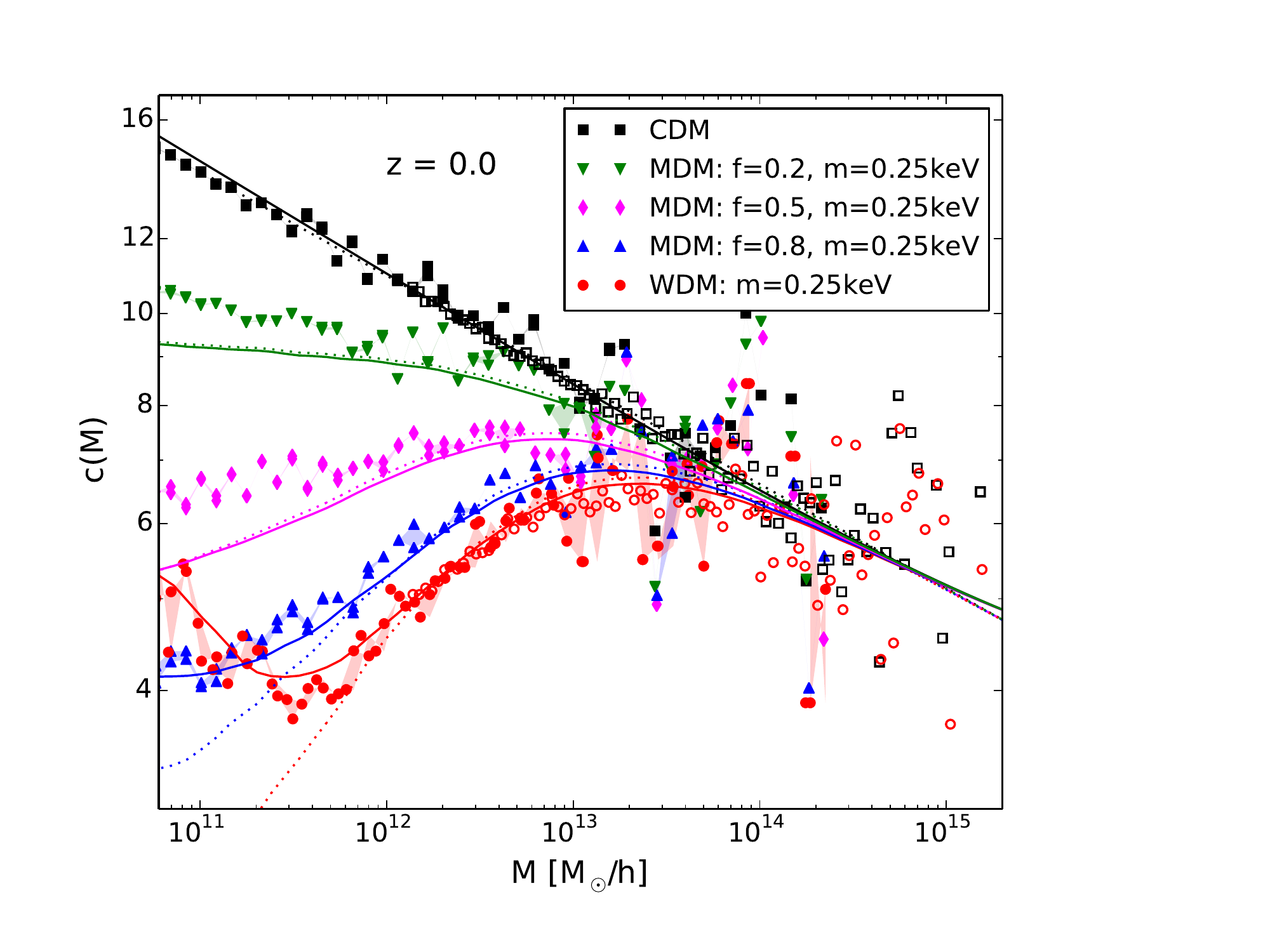}
\includegraphics[trim = 12mm 4mm 42mm 14mm, clip, scale=0.44]{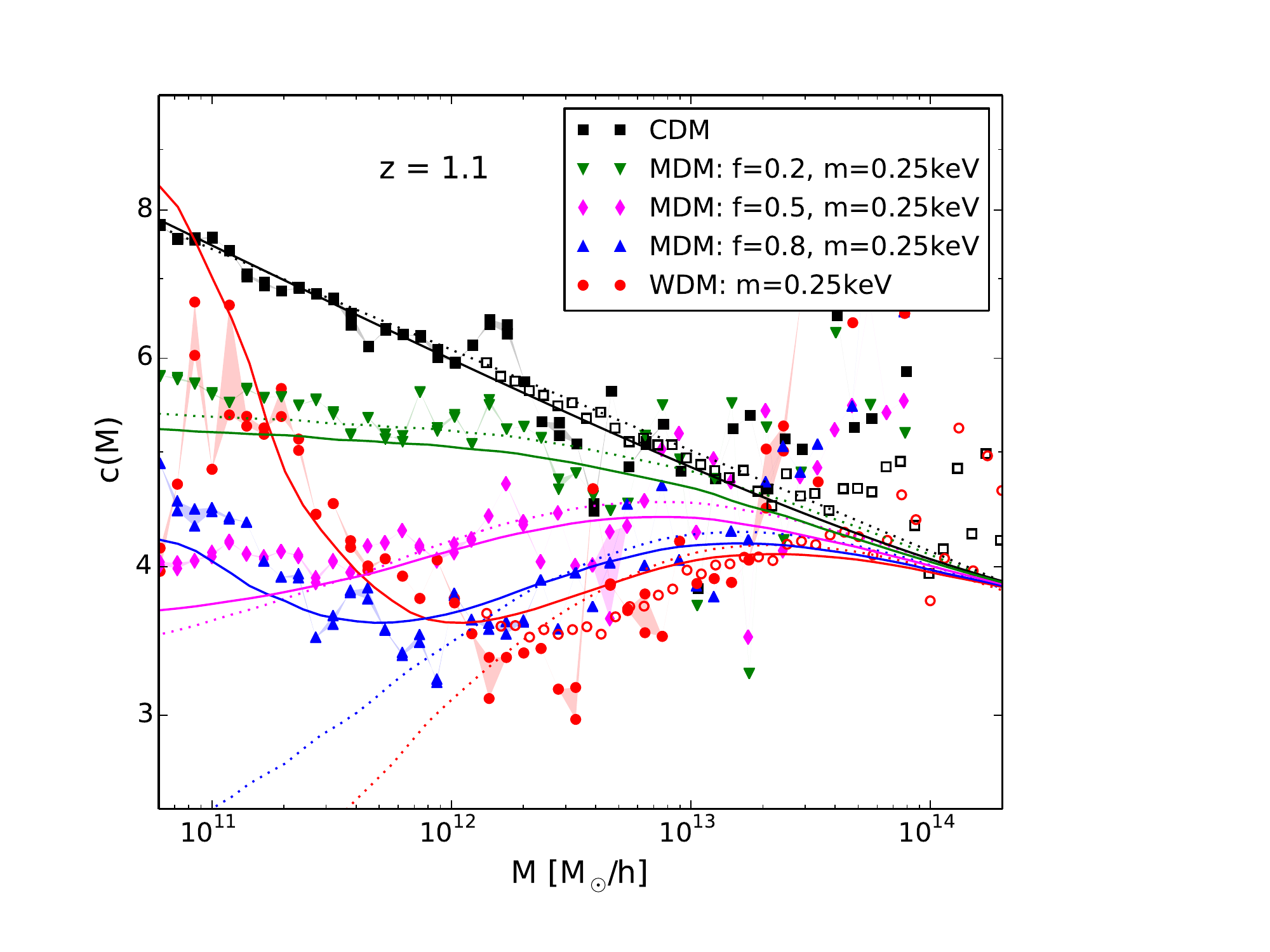}
\caption{Concentration-mass relation for CDM (black), different WDM (red, cyan, purple) and different MDM models (blue, magenta, green) with circumjacent shaded regions representing the uncertainty due to artefact subtraction. Symbols illustrate the measurements from simulations, solid and dotted lines correspond to the models presented in the text (based on CDM fitting functions with and without upturn on cluster scales).}\label{fig:concentrations}
\end{figure*}

In Fig. \ref{fig:z_collapse} the average collapse redshift from Eq. (\ref{Dcollapse}) is plotted against halo mass at redshift zero. Here we used the fraction $F=0.05$ and the sharp-$k$ filter to calculate $S_\chi$, but similar results are obtained with a tophat filter if the fraction is changed to $F=0.01$. For CDM (black line) $z_c$ increases monotonically towards smaller masses, while for WDM (red line) and MDM (blue, magenta, and green line) it increases, turns over, and decreases again. Very small and very large WDM and MDM haloes are thus rather young, while medium sized haloes are considerably older. It is important to note that this peculiar behaviour is no sign for the failure, but rather a natural prediction of hierarchical structure formation: suppressed perturbations can only collapse if they sit on top of larger perturbations, but this means that they only live for a short time before they get cannibalised by the collapsing larger perturbation.

As mentioned earlier, Eq.~(\ref{Dcollapse}) provides the average collapse redshift of all progenitors and is therefore expected to under-predict the actual collapse redshift of the halo. In Fig. \ref{fig:z_collapse} we show that this is indeed the case: the average collapse redshift measured in CDM simulations \citep[solid grey stars, obtained from \citet{Jiang2014}, based on simulations from][]{Zhao2009} is about a factor of 1.5 larger than the one predicted by our simplified model. It is, however, interesting to realise that the slope of the CDM $z_c$-mass relation described by Eq. (\ref{Dcollapse}) agrees well with the one from simulations. This is illustrated by the empty grey stars which are the same data normalised to the black line. In the next section we will see that this relative agreement is sufficient to derive a physically motivated recipe for the concentration-mass relation.

\subsection{Concentrations}
Theory-based arguments can be used to understand the qualitative behaviour of the concentration-mass relation, but there is no consistent theoretical model that provides a prediction. It is therefore common practice to measure concentrations in simulations and to use fits for further investigations. Here we present a simple method which is based on prior knowledge of pure CDM concentrations and predicts the concentration-mass relation for any scenario with suppressed small-scale perturbations.

Let us suppose we know a function describing the concentration-mass relation for the CDM case. In this paper we use two different fitting functions: the first corresponds to a simple power-law proposed by \citet{Maccio2008}
\begin{equation}\label{c_CDM0}
c_{\rm CDM}(M)=\alpha\left(\frac{10^{12} h^{-1}\rm M_{\odot}}{M}\right)^{\gamma},
\end{equation}
the second includes an additional parameter $\mathcal{M}$ accounting for an increase at large masses
\begin{equation}\label{c_CDM1}
c_{\rm CDM}(M)=\alpha\left(\frac{10^{12}h^{-1}\rm M_{\odot}}{M}\right)^{\gamma}\left[1+\left(\frac{M}{\mathcal{M}}\right)^{0.4}\right],
\end{equation}
as described by \citet{Klypin2014}. Based on one of these fits, it is possible to obtain a concentration-mass relation for any DM scenario $\chi$ by simply assuming that haloes with the same collapse redshifts end up having the same concentrations. In other words, we can use Eq.~(\ref{Dcollapse}) to equate
\begin{equation}
\langle D\rangle_{\rm CDM}(M_{\rm CDM})\equiv \langle D\rangle_{\chi}(M_{\chi}),
\end{equation}
and determine a function $M_{\rm CDM}(M_{\chi})$ relating the halo mass of any DM scenario $\chi$ to the corresponding halo mass of CDM. This function can then be used in either Eq. (\ref{c_CDM0}) or Eq. (\ref{c_CDM1}) to obtain the concentration-mass relation of scenario $\chi$, i.e
\begin{equation}\label{c_chi}
c_{\chi}(M_{\chi})=c_{\rm CDM}[M_{\rm CDM}(M_{\chi})].
\end{equation}
The procedure explained here is physically motivated, as it simply assigns the same value for the concentration to haloes with the same collapse redshift, independently of the DM scenario.

In Fig.~\ref{fig:concentrations} we compare the concentrations from simulations (filled and empty symbols) to the model described by Eq. (\ref{c_chi}) utilising the fitting functions from Eq.~(\ref{c_CDM0}) (dotted lines) and Eq.~(\ref{c_CDM1}) (solid lines). The left panel shows the concentrations of CDM (black) and WDM (cyan, purple) at redshift 0. While the concentration-mass relation of the CDM model exhibits the expected monotonic increase towards small masses, there is a clear turn-over visible in the WDM scenarios. The model given by Eq.~(\ref{c_chi}) is able to reproduce this behaviour for either of the underlying fitting functions (using the parameters $\alpha=10.84$, $\gamma=0.085$, and $\mathcal{M}=5.5\times10^{17}$ $h^{-1}\rm M_{\odot}$).

The middle panel of Fig.~\ref{fig:concentrations} shows concentrations of CDM (black), WDM (red), and MDM (blue, magenta, green) at redshift 0. As expected, the CDM concentrations increase monotonically towards small masses, while the WDM concentrations exhibit a turnover. For the case of MDM, the concentration-mass relation is either flattened compared to CDM (green dots) or it turns over as in the WDM scenario (magenta and blue dots) depending on the mixing fraction. Additionally to this general trend, the red and blue dots (representing the most extreme cases in terms of suppression) exhibit a second upturn at small masses below $10^{12}$ $h^{-1}\rm M_{\odot}$. This feature can be understood by comparing the collapse redshifts:  haloes with masses around this secondary upturn have very low collapse redshifts comparable to the ones from the largest haloes with masses above $10^{15}$ $h^{-1}\rm M_{\odot}$. As a result, their concentrations should coincide, and an increase of the concentration-mass relation towards very large masses directly translates into an upturn towards small masses around the suppression-scale. This is the reason why the model based on the fit with large-scale upturn (Eq.~\ref{c_CDM1}) is in good agreement with the simulations, while the model based on the simple power-law fit (Eq.~\ref{c_CDM0}) fails at the smallest mass scales\footnote{The origin of the upturn of concentrations towards very high halo masses is disputed: while \citet{Ludlow2014} argue that the largest haloes are not fully virialised, \citet{Klypin2014} claim that many large haloes grow out of unusually spherical perturbations, leading to a boost of the concentrations.}. For the fitting parameters we used $\alpha=10.96$, $\gamma=0.12$, and $\mathcal{M}=5.5\times10^{17}$ $h^{-1}\rm M_{\odot}$ in the former and $\alpha=10.96$ and $\gamma=0.11$ in the latter case. These values are very close to results from \citet{Klypin2014} and \citet{Dutton2014}, respectively, except for a small change of normalisation due to different halo mass definitions.

In the right panel of Fig.~\ref{fig:concentrations}, the same models are plotted at redshift 1.1. As a general trend, the concentration-mass relations are flatter than at redshift 0 which is in agreement with former studies \citep{Prada2012,Dutton2014}. The secondary upturn towards small masses is, however, more prominent than at redshift 0. This is a direct consequence of the fact that the large-scale upturn of pure CDM concentrations shifts towards smaller masses if higher redshifts are considered  \citep{Klypin2014}. In this panel we used the parameters $\alpha=5.9$, $\gamma=0.1$, and $\mathcal{M}=4.4\times10^{16}$ $h^{-1}\rm M_{\odot}$ for the fit with large-scale upturn and $\alpha=6.1$, $\gamma=0.085$ for the simple power-law fit.

In summary, the model based on Eq.~(\ref{c_chi}) is able to reproduce the concentration-mass relation of scenarios with various small-scale power suppressions provided the concentrations of pure CDM are well known. In agreement with previous papers, we observe that concentrations seem to be tightly connected to the halo collapse redshifts. In the presence of a strong suppression of perturbations (such as in WDM or some MDM scenarios), this may lead to a wave-like feature in the concentration-mass relation, reflecting the low collapse redshifts of haloes around the suppression scale.


\section{Conclusions}\label{sec:conclusions}
The standard model of cosmology predicts a suppression of perturbations below a characteristic scale depending on the dark matter (DM) candidate. For the case of DM consisting of weakly interactive masseuse particles (WIMP), the suppression scale is orders of magnitudes below the range of current astronomical observations, and only a potential detection of the WIMP annihilation signal could eventually change this. For alternative DM candidates, such as the sterile neutrino, the suppression scale is expected to be large enough to have an effect on current galaxy observations. In order to distinguish between different DM scenarios, it is therefore crucial to understand nonlinear structure formation around the scale of suppressed perturbations.

In this paper we study structure formation of various suppressed initial power spectra using both numerical simulations and analytical techniques. The simulations are deliberately chosen to cover a variety of different scenarios from steep cutoffs to shallow suppressions occurring on a wide range of scales. We study warm and WIMP DM both exhibiting steep cutoffs as well as mixed DM with much shallower suppressions depending on the mixing fraction.

Many of the simulations performed here suffer from artificial clumping, a well known problem in WDM simulations \citep{Goetz2003, Wang2007, Elahi2014,Reed2014}. We therefore apply a post-processing scheme similar to the one presented by \citet{Lovell2014}, which filters out artefacts based on both proto-halo sphericity and overlap between two simulations with the same density field but different resolution.

Along with the simulations, we develop analytically motivated models for the halo and subhalo mass functions, the halo collapse redshift, and the concentration-mass relation which are designed to work for arbitrary small-scale suppressions. The main findings of the paper are summarised in the following list:
\begin{itemize}
\item We further investigate the extended Press-Schechter model based on the sharp-$k$ filter function introduced by \citet{Schneider2013}. We show that it provides a simple prescription of the halo mass function, matching the simulation outcomes for all models studied here. This includes various warm DM, mixed DM, WIMP-DM, and pure cold DM scenarios, covering suppression scales at very different scales and with a variety of different shapes. It is therefore fair to assume that the sharp-$k$ model can be used to predict the halo abundance for cosmologies with arbitrary small-scale suppressions.
\item We present a method to estimate the number of substructures in a host halo, and show that it agrees very well with warm DM simulations from \citet{Lovell2014}. Based on this, we provide approximative constraints on mixed DM models by comparing the expected number of substructures to the observed number of Milky-Way satellites.
\item Using the conditional mass function, we determine the average collapse redshift of progenitors. Normalising this relation to measurements from CDM simulations yields a simple estimation for the collapse redshifts of a halo in any DM scenario. Well above the suppression scale, smaller haloes tend to be older than larger ones. This is not the case around the suppression scale where haloes are very young on average. We argue that, despite this turnaround in the $z_c$-mass relation for models with suppressed perturbations, structure formation remains a hierarchical process in the sense that there is now sign of halo-formation via fragmentation.
\item Based on the collapse redshifts of haloes, we develop a recipe for the concentration-mass relation in the presence of arbitrary suppressed small-scale perturbations and show that it agrees surprisingly well with simulations. For the case of warm DM and some mixed DM scenarios, the concentration-mass relation decreases towards smaller masses when approaching the suppression scale. At even smaller scales, however, the relation grows again, an effect tightly connected to the recently observed increase of concentrations at the largest scales in pure CDM simulations \citep{Prada2012,Klypin2014}.
\end{itemize}

The findings presented in this paper are potentially useful for analytical and semi-analytical modelling of the nonlinear Universe. In the presence of suppressed small-scale perturbations, analytical methods are particularly justified because numerical simulations notoriously suffer from artefacts and need to be analysed with great care.

\section*{Acknoledgments}
I would like to thank Darren Reed and J\"uerg Diemand for many helpful discussions related to this work. Furthermore, I'm grateful to Doug Potter and Alex Knebe for invaluable support with various computational issues. The simulations have been performed on the following computer clusters: z-box (University of Zurich), Piz Daint (Swiss National Super-Computer Centre, CSCS), and Lonestar (Texas Advanced Computer Center). This work is supported by the Swiss National Science Foundation through the early researcher fellowship (P2ZHP2\_151605).



\appendix
\section{Some more about removing artefacts}\label{sec:appendix}

\begin{figure*}
\includegraphics[trim = 6mm 4mm 20mm 10mm, clip, scale=0.23]{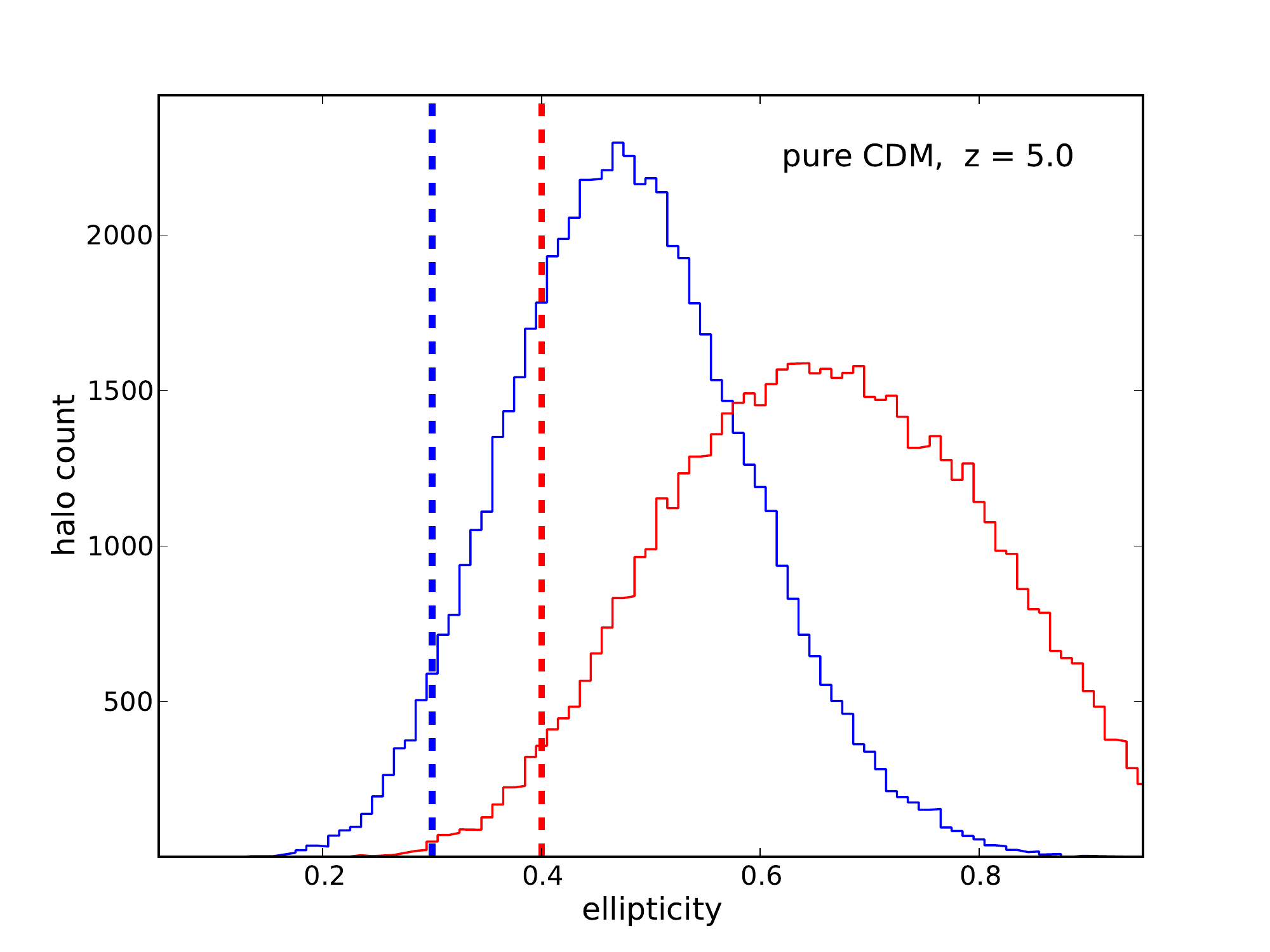}
\includegraphics[trim = 6mm 4mm 20mm 10mm, clip, scale=0.23]{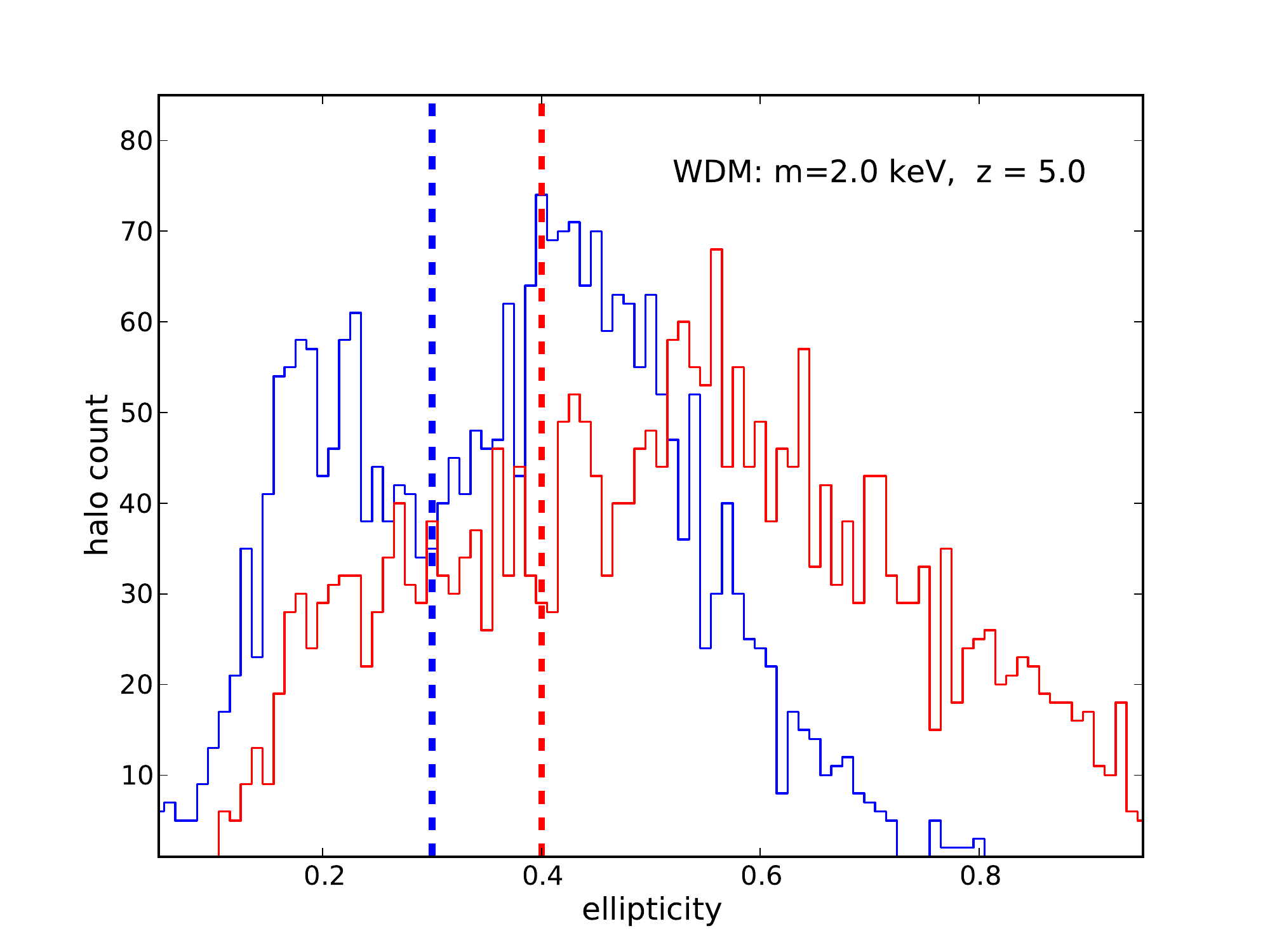}
\includegraphics[trim = 6mm 4mm 20mm 10mm, clip, scale=0.23]{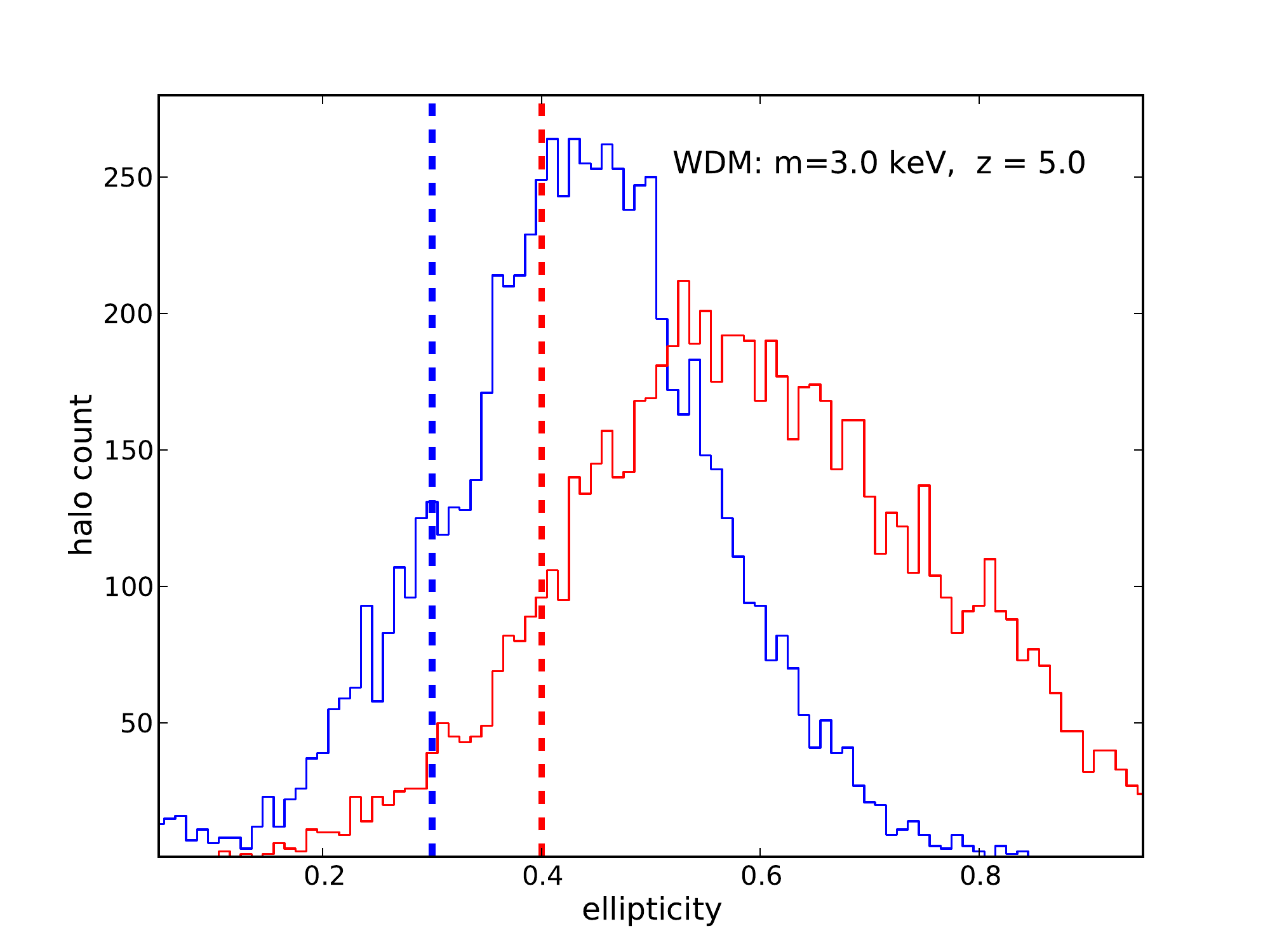}
\includegraphics[trim = 6mm 4mm 20mm 10mm, clip, scale=0.23]{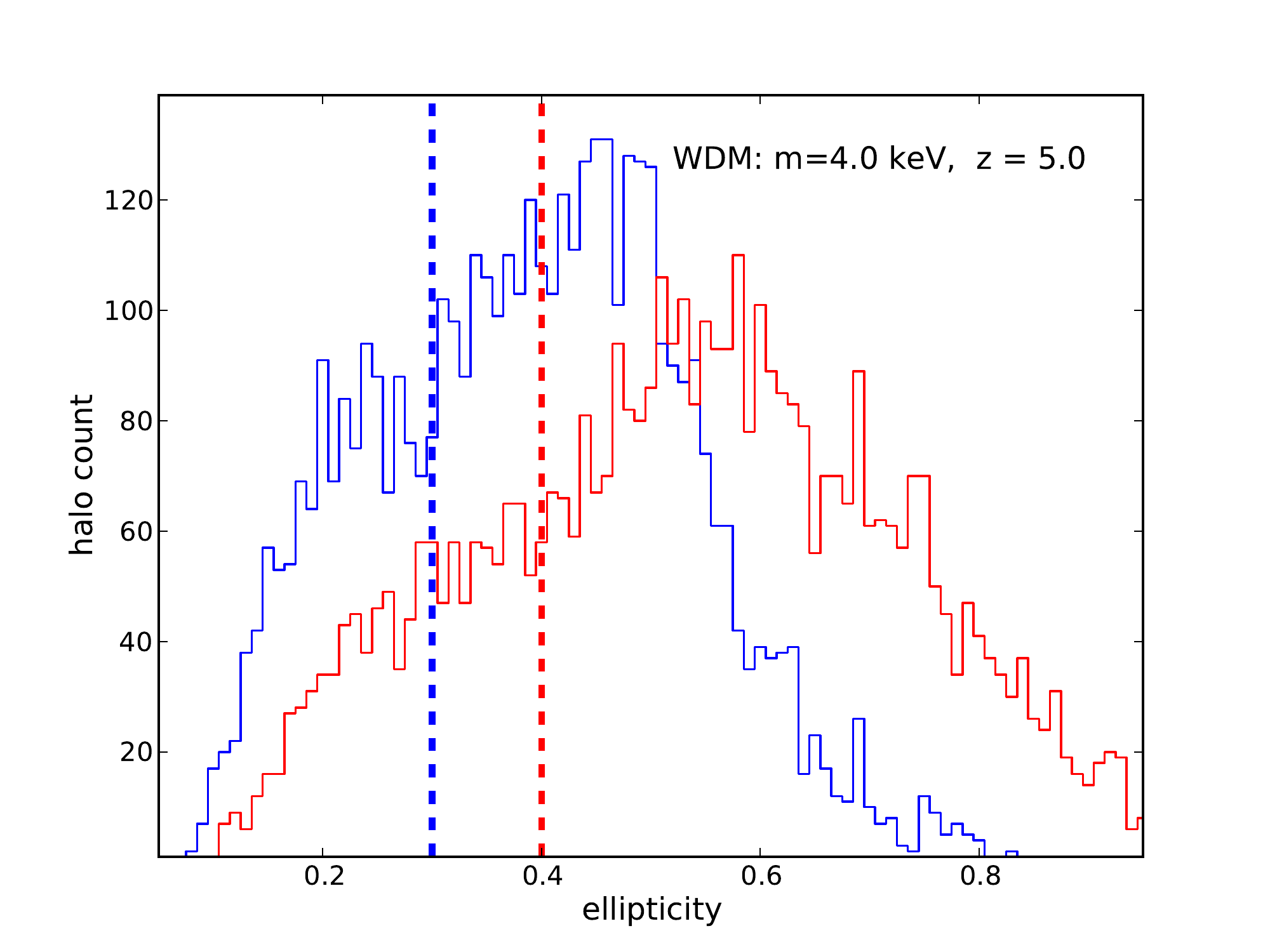}\\
\includegraphics[trim = 6mm 4mm 20mm 10mm, clip, scale=0.23]{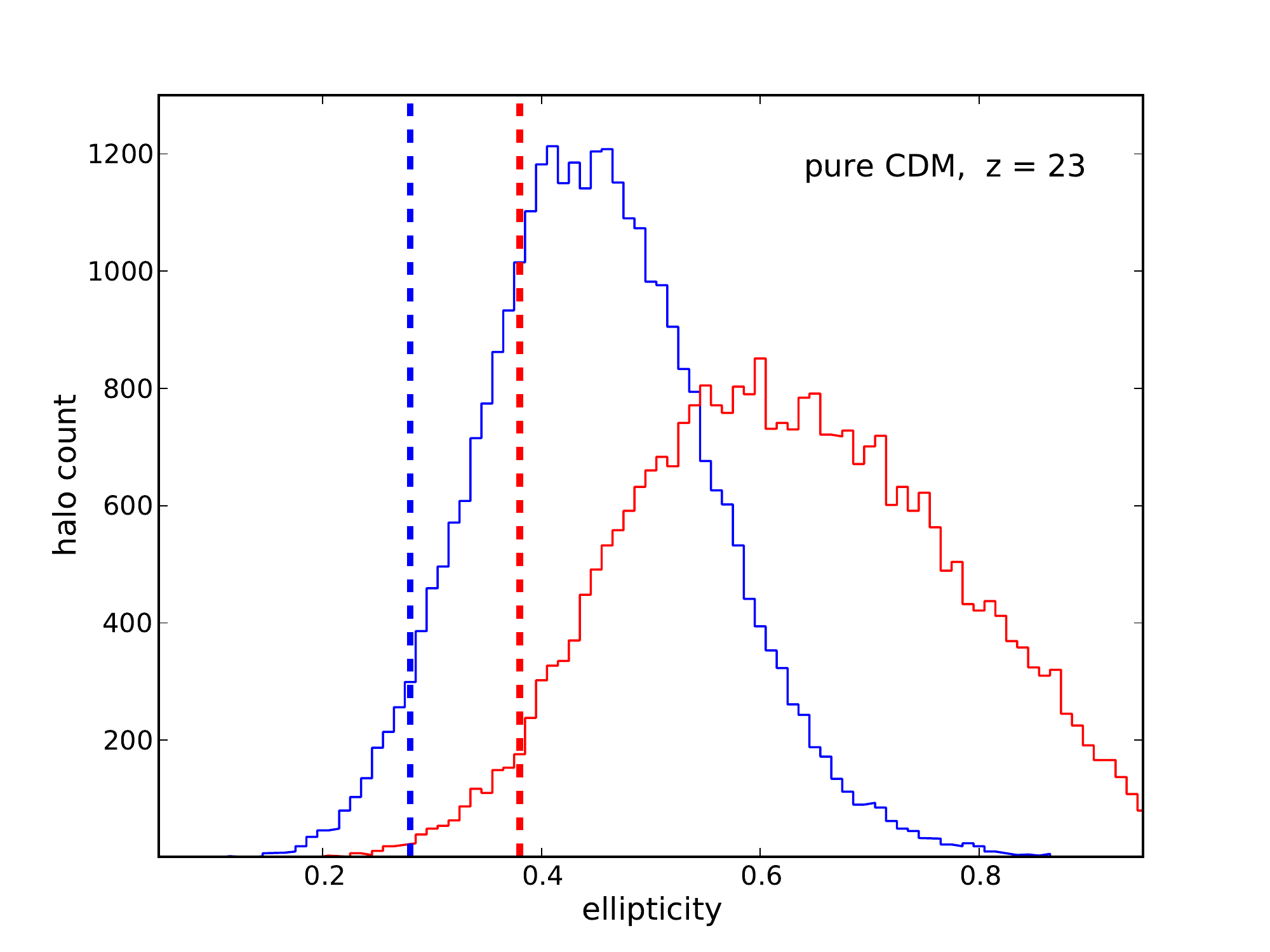}
\includegraphics[trim = 6mm 4mm 20mm 10mm, clip, scale=0.23]{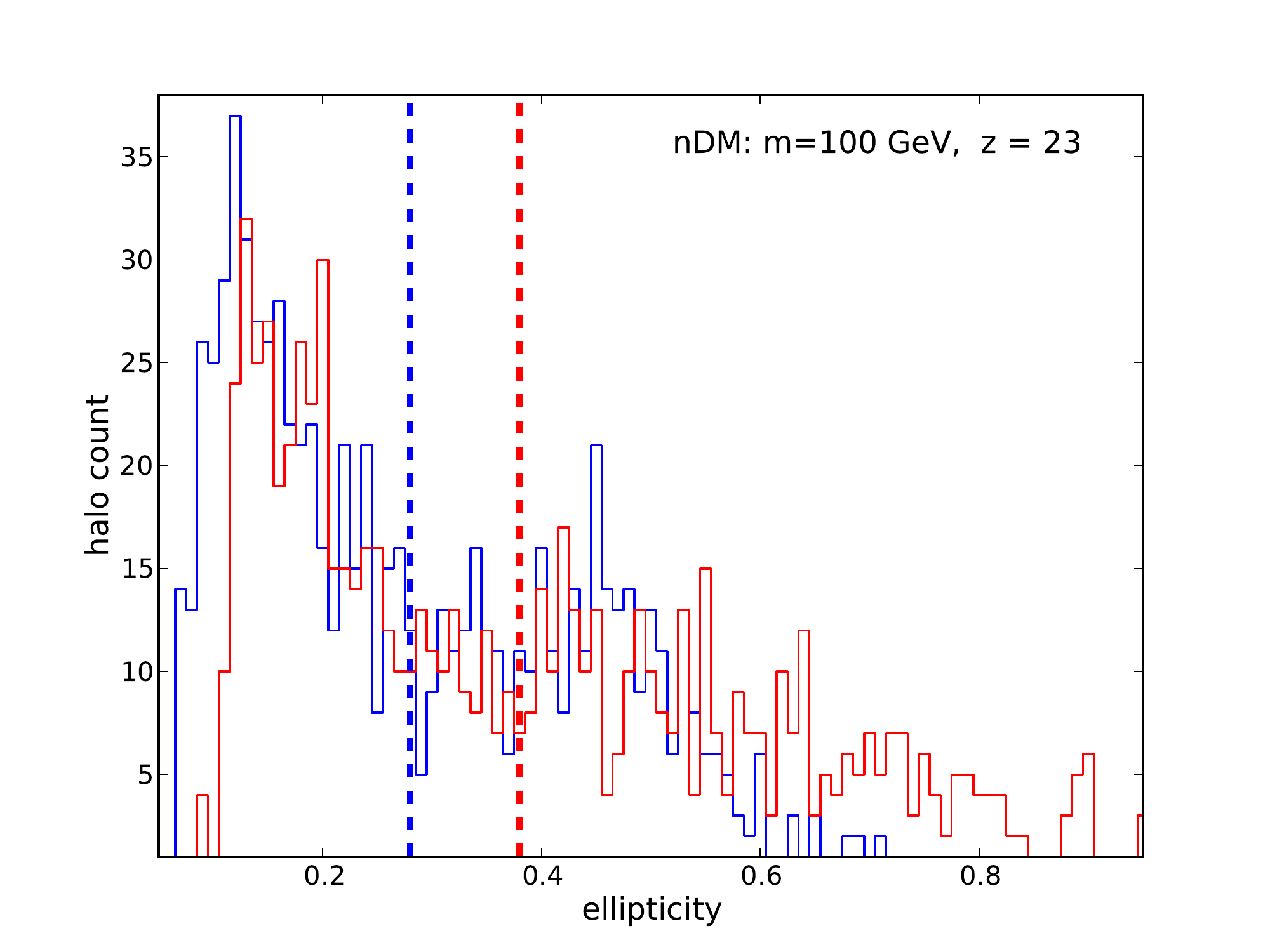}
\includegraphics[trim = 6mm 4mm -162mm 10mm, clip, scale=0.23]{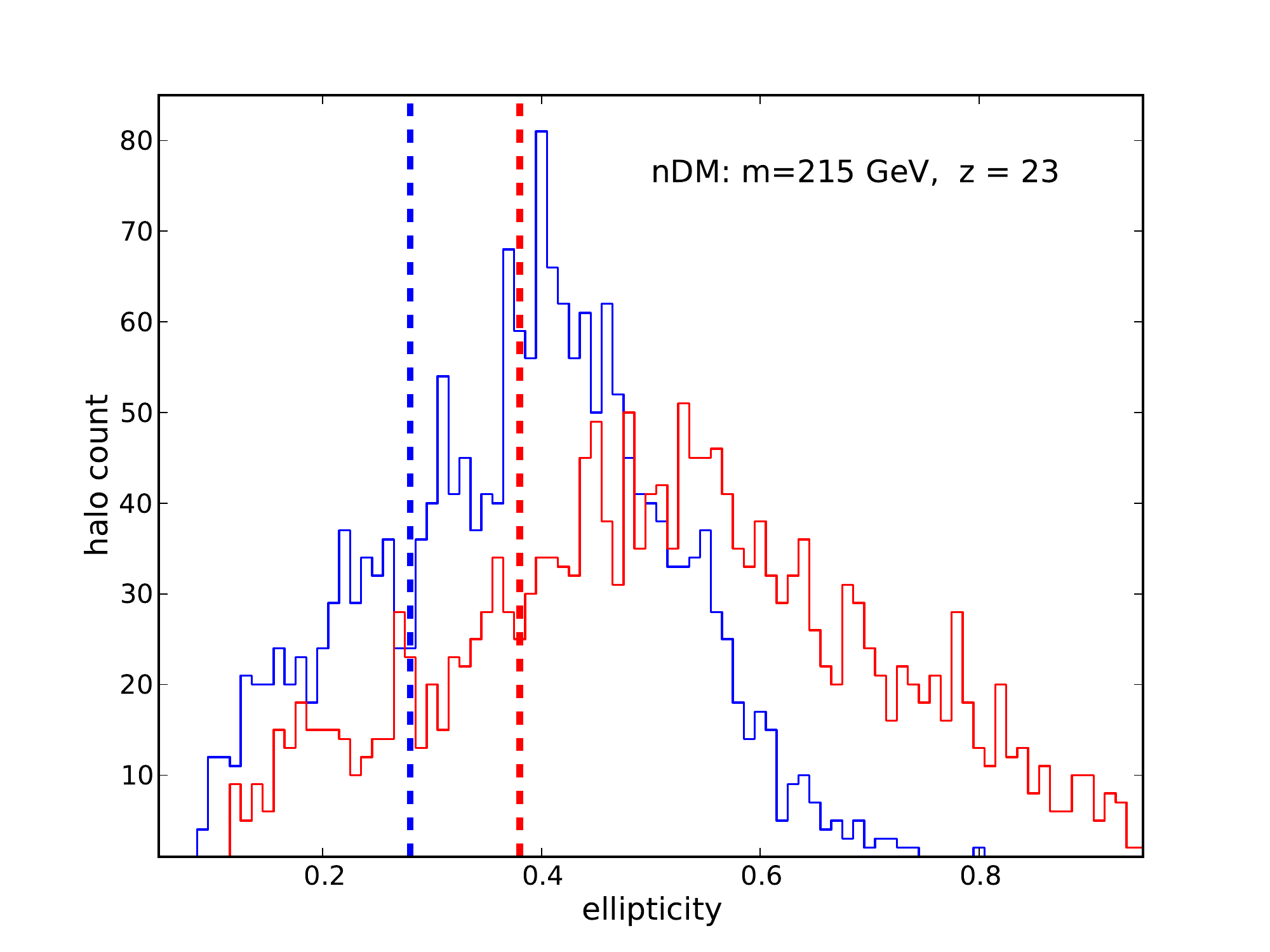}
\caption{Histograms of the halo ellipticity parameters $s$ (blue) and $q$ (red) not shown in the main text. Top: CDM and WDM with $m=2, 3, 4$ keV (from left to right) at redshift 5. Bottom: pure CDM and WIMP DM with neutrino masses of $m=100, 215$ GeV (from left to right) at redshift 23. Vertical dashed lines illustrate the {\it sphericity cut}.}
\label{fig:Appendixhalohistogram}
\end{figure*}
\begin{figure*}
\includegraphics[trim = 8mm 4mm 40mm 14mm, clip, scale=0.55]{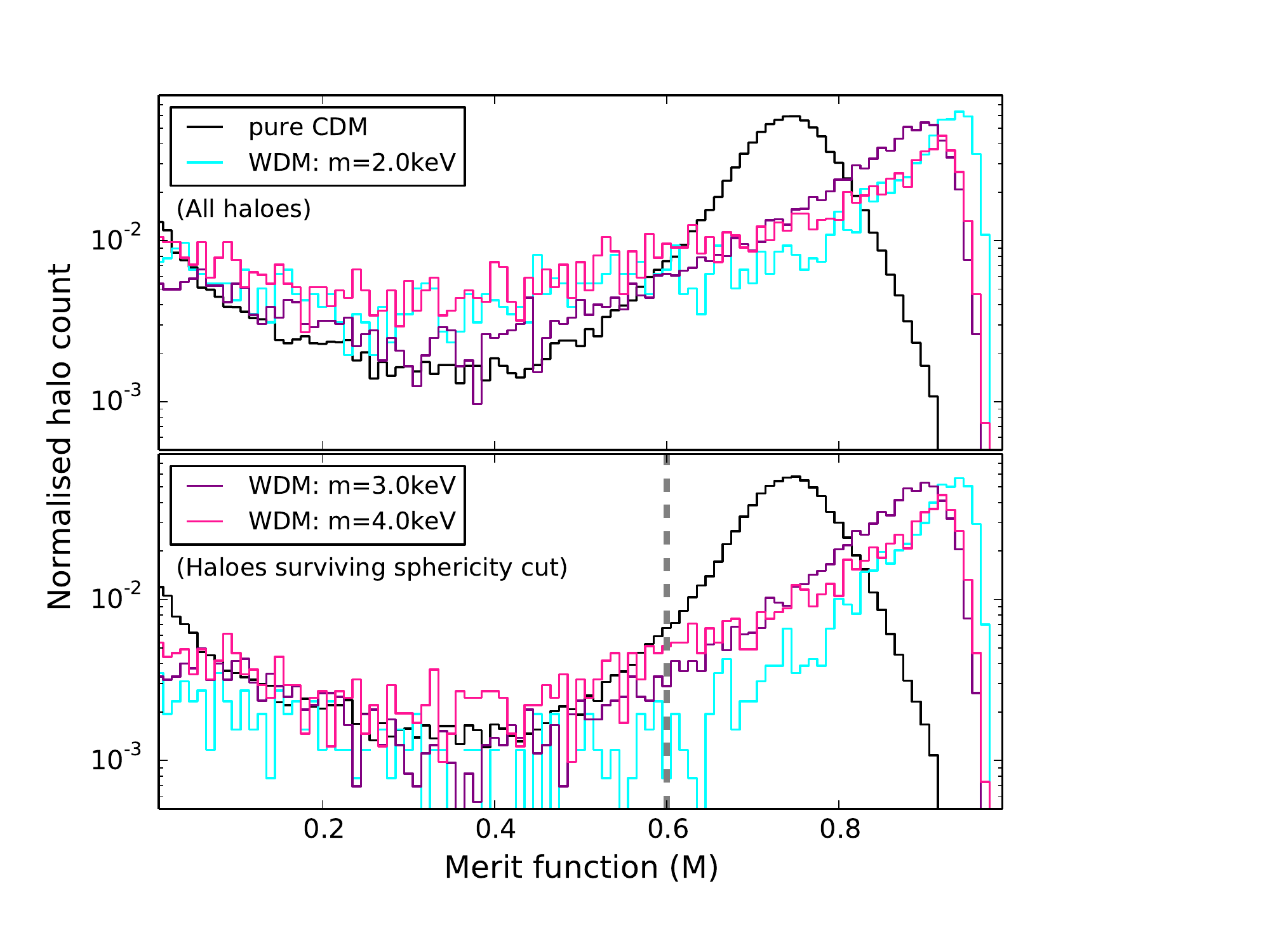}\includegraphics[trim = 8mm 4mm 40mm 14mm, clip, scale=0.55]{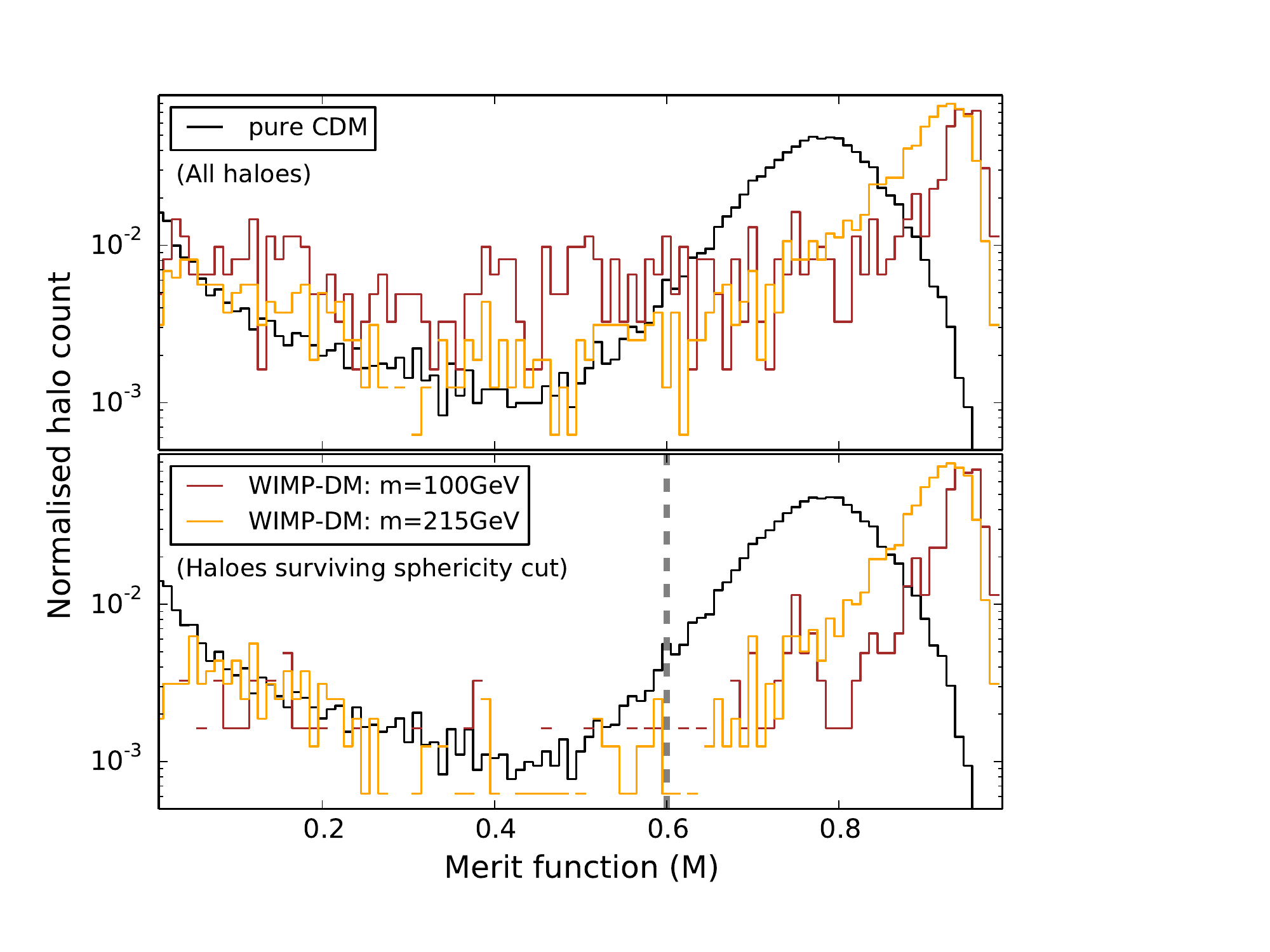}
\caption{Histograms of the merit function $M$ (i.e.Eq.~\ref{merit}) of all haloes before (top) and after (bottom) the {\it sphericity cut}. Left: CDM (black) and WDM with $m=2, 3, 4$ keV (cyan, magenta, pink) at redshift 5. Right: pure CDM (black) and WIMP-DM with neutrino masses of $m=100, 215$ GeV (brown, orange) at redshift 23. Vertical dashed lines correspond to the {\it resolution cut}.}
\label{fig:Appendixremaininghalohistogram}
\end{figure*}

In the main text we have discussed how the removal of artefacts affects the halo samples of the CDM, WDM ($m=0.25$ keV) and MDM simulations at redshift zero. Here we focus on the remaining DM models studied in this paper, providing the details of the CDM and WDM ($m=2,3,4$ keV) simulations at redshift 5 as well as the pure CDM and WIMP DM models at redshift 23.

The distinction between real haloes and artefacts is somewhat easier towards high redshifts, since most of the haloes are in a less evolved state. This allows us to choose higher values for the sphericity parameters plus a less restrictive minimal number of 500 particles per halo. 

For the CDM and WDM models (with $m=2,3,4$ keV) at redshift 5, we use $s_{\rm c}=0.3$ and $q_{\rm c}=0.4$ for the {\it sphericity cut}, which are values that filter out a maximum of WDM artefacts while removing less than 5 percent of all CDM haloes. The distributions of sphericity parameters $s$ (blue) and $q$ (red) are shown in the top panels of Fig.~\ref{fig:Appendixhalohistogram}. The distributions of the WDM model with $m=2$ keV is characterised by two peaks, while the other WDM models show broadened distributions compared to CDM, indicating the presence of artificial haloes\footnote{The WDM model with $m=4$ keV is run with a smaller box size than the other WDM models of the series, a fact that leads to a further broadening of the histogram.}. The distribution of the merit function (quantifying the overlap of haloes from different resolution) is illustrated in the left-hand-side of Fig.~\ref{fig:Appendixremaininghalohistogram}. The top panel includes all haloes (before the {\it sphericity cut}), clearly showing the excess of artefacts in the WDM simulations at $M<0.6$. This excess is greatly reduced after the {\it sphericity cut}, as illustrated in the bottom panel of Fig.~\ref{fig:Appendixremaininghalohistogram}. Further artefacts are removed with the {\it resolution cut} $M_{\rm c}=0.6$, illustrated as vertical dashed line in the lower panel of Fig.~\ref{fig:Appendixremaininghalohistogram}.

For the models of pure CDM and WIMP DM (with neutralino mass of $m=100$ and $m=215$ GeV) at redshift 23, we apply a {\it sphericity cut} of $s_{\rm c}=0.28$ and $q_{\rm c}=0.38$, which again removes a maximum of artefacts while affecting pure CDM by less than 5 percent. The sphericity distributions of these DM models are presented in the bottom panels of Fig.~\ref{fig:Appendixhalohistogram}, showing the usual picture of a Gaussian distribution for pure CDM and much broader, occasionally double-peaked distributions for WIMP DM. The distributions of the merit function are illustrated on the right-hand-side of Fig.~\ref{fig:Appendixremaininghalohistogram}, the top and bottom panel showing the cases before and after the {\it sphericity cut}. We again observe an excess of haloes in the WIMP DM models at low $M$ with respect to pure CDM which is greatly reduced once the {\it sphericity cut} is applied. The {\it resolution cut} $M_{\rm c}=0.6$ is illustrated as vertical dashed line in the bottom panel of Fig.~\ref{fig:Appendixremaininghalohistogram}.

\end{document}